\documentclass[journal=jpcafh,manuscript=article,layout=onecolumn]{achemso}
\usepackage{amsthm,amsmath,amsfonts,amssymb,graphicx}
\usepackage{graphicx}
\usepackage{subfigure}
\usepackage{color}  
\usepackage{multirow}
\usepackage{ifthen}
\newboolean{twocolumn}
\ifthenelse{\lengthtest{\columnwidth < 300pt}}
     {\setboolean{twocolumn}{true}}     %% some equations split over two lines, figures and tables resized or repositioned
     {\setboolean{twocolumn}{false}}    

\author{T. J. Price}
\affiliation{Department of Physics and Astronomy, Purdue University, West Lafayette, Indiana 47907, USA}
\email{price198@purdue.edu}
\author{Chris H. Greene}
\affiliation{Department of Physics and Astronomy, Purdue University, West Lafayette, Indiana 47907, USA}
\alsoaffiliation{Purdue Quantum Center, Purdue University, West Lafayette, Indiana 47907, USA}
\email{chgreene@purdue.edu}

\title{Semiclassical Treatment of High-Lying Electronic States of H$_2^+$}
\abbreviations{PEC,DME,HJ}

\begin{document}
\begin{tocentry}
\includegraphics[width=2.2in]{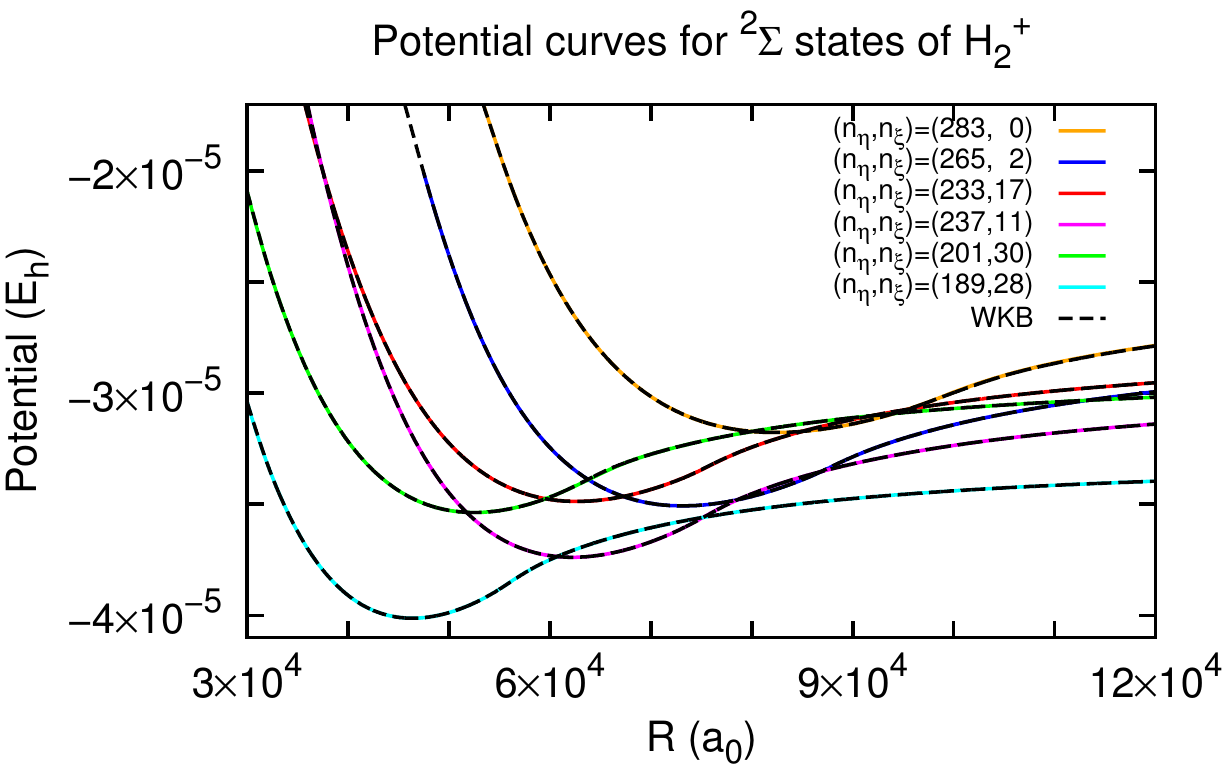}
\end{tocentry}

\begin{abstract}
This work reports quantum mechanical and semiclassical WKB calculations for energies and wave functions of high-lying $^2\Sigma$ states of H$_2^+$ in atomic units. The high-lying states presented lie in an unexplored regime, corresponding asymptotically to H $(n\leq 145)$ plus a proton, with $R\leq 120,000$~$a_0$. We compare quantum mechanical energies, spectroscopic constants, dipole matrix elements, and phases with semiclassical results and demonstrate a good level of agreement. The quantum mechanical phases are determined by using Milne's phase-amplitude procedure. Our semiclassical energies for low-lying states are compared with those published previously in the literature.  
\end{abstract}

\section{Introduction}
Many calculations for energies and wave functions of H$_2^+$ have been performed, one before quantum mechanics was invented \cite{Pauli}. Some have involved simplified models \cite{Frost,Frost2}, while others utilize the separability of H$_2^+$ in spheroidal coordinates to obtain accurate, almost exact results (see e.g. \cite{Jaffe,Baber,Bates,Hodge,Montgomery,Bailey,Madsen,Boyack,TonyScott,Leaver,Liu,Novello,Figueiredo,Ponomarev,Ramaker,Ramaker2,Defrance,Defrance2,Somov,YooandGreene,Santos,Zammit,Larsson} and the references therein). 

To our knowledge, the highest energy calculations were carried out by Pelisoli, Santos, and Kepler \cite{Santos,Santos2} for internuclear separations up to 300 $a_0$ and dissociation limits with principal quantum numbers $n\leq 10$, to study spectral line broadening of hydrogen in the atmospheres of white dwarf stars and laboratory plasmas; there has been discussion in the literature regarding the lack of state-resolved data for H$_2^+$ \cite{Zammit,Babb}. In this paper, we include quantum calculations for higher excited electronic states which have not yet been calculated.

Our interest in highly excited states of H$_2^+$ was sparked by the desire to accurately calculate long-range potential energy curves (PECs) for Rydberg molecules. A promising approach, (generalized) local frame transformation theory \cite{GLFT,LFT,LFT2}, would involve such states of H$_2^+$. The essence of the theory is to identify systems with two regions of different approximate symmetry and to interrelate the wave functions in each region. One very successful application has been to the photoionization of non-hydrogenic Rydberg atoms in external electric fields. In that application, hydrogenic states are used as a starting point. For some long-range homonuclear Rydberg molecules, the starting point would be states of H$_2^+$.

The states are so numerous, however, that an efficient approximation scheme is warranted. For an approximate treatment of long-range PECs of H$_2^+$, several approaches could be taken. One might consider using Rayleigh-Schr\"odinger perturbation theory for a hydrogen atom perturbed by a proton. Although it has been shown that accurate long-range energies of H$_2^+$ cannot be obtained with RSPT alone \cite{DalgarnoLewis,Robinson}, semiclassical \cite{Cizek} and other \cite{Tang,Scott} methods have been introduced to treat the exponentially small exchange terms. Symmetry adapted perturbation expansions have been introduced as well\cite{Szalewicz}. 

On the other hand, semiclassical, WKB-like methods for energies and separation constants of H$_2^+$ have also led to good agreement with quantum results \cite{Reinhardt,Gershtein,Froman1,Froman2,Sink,Hunter,Hnatic,Hnatic2,Hnatic3,Guaranho}, and we chose to take that approach. Existing semiclassical treatments of H$_2^+$ are either carried out for the general two-center Coulomb case, or are refined for accurate treatment of the lowest-lying states, or both. The analyses can be complicated because of the presence of a potential barrier between the two nuclei. Some authors go around the classical turning points in the complex plane to derive quantization conditions that include reflection just above the barrier \cite{Gershtein,Froman1,Froman2,Hunter}, while others utilize uniform approaches to derive quantization conditions that account for the coalescence of the classical turning points at the barrier top \cite{Reinhardt,Sink,Hnatic,Hnatic2,Hnatic3,Guaranho}. While these refinements are essential for the lowest-lying states of H$_2^+$, for higher states, only very small portions of the PECs involve energies near the height of the barrier, and a simple treatment of the system can supplant more refined methods with little loss of accuracy. 

We derive quantization conditions by a simple WKB treatment, with tunneling taken into account, drawing on some results from the literature, namely a Langer-type \cite{Langer} modification for H$_2^+$ \cite{Gershtein} and the connection formula for a first-order pole  as applied to H$_2^+$ \cite{Froman1}. We compare our WKB energies with those of Strand and Reinhardt \cite{Reinhardt} and Gershtein \latin{et al.}\ \cite{Gershtein}, and confirm that for low-lying states our approach leads to insufficient accuracy because the top of the barrier is in an important region of the potential. For higher states, however, very good accuracy is achieved with our quantization conditions; the $\nu$ values agree to three or four decimal places except for in small regions of each potential curve near the barrier top. 

In this work, we also compare quantum and semiclassical wave functions; to our knowledge such a comparison has not been previously reported. Dipole matrix elements with the ground state are presented, along with the phases of the wave functions. For the phase comparisons, we implement the Milne phase-amplitude procedure \cite{Milne} as applied to H$_2^+$ \cite{Larsson}, with the Milne equation expressed as a third order, linear differential equation \cite{CoteMilne,Kiyokawa}.

Since we will compare with their results, we now briefly summarize the approaches of Strand and Reinhardt \cite{Reinhardt} and Gershtein \latin{et al.}\ \cite{Gershtein}.

Strand and Reinhardt \cite{Reinhardt} have derived quantization conditions and wave functions using both a primitive and uniform WKB approach. In particular, Strand and Reinhardt have emphasized the non-uniqueness of the variable transformations which lead to {modified} quasimomenta. In order to obtain canonically invariant amplitudes and phases, they have employed canonical transformations and the Keller-Maslov quantization conditions, which assume a single classical allowed region, to determine primitive WKB solutions; as such, these results do not incorporate tunneling through the barrier. To incorporate tunneling effects, Strand and Reinhardt use the parabolic uniform approximation \cite{Miller}, where parabolic cylinder functions are used as comparison functions rather than exponentials. 

Strand and Reinhardt determine wave functions that have the same form as those given by Gershtein \latin{et al.}\ \cite{Gershtein}. Gershtein \latin{et al.}\ used modified, in this case referred to as Bethe-modified, ``quasimomenta'' that agree with the momenta in the {separated} Hamilton-Jacobi (HJ) equation. Using these forms of the quasimomenta, Gershtein \latin{et al.}\ treated the system using the ``complex method''\cite{Furry,Heading,Kemble}, which was invented by Zwaan and involves going around the classical turning points in the complex plane. 

\section{Theoretical Methods}
\subsection{Quantum Treatment}
\begin{figure}[!hb]
\centering
\includegraphics[width=2.5in]{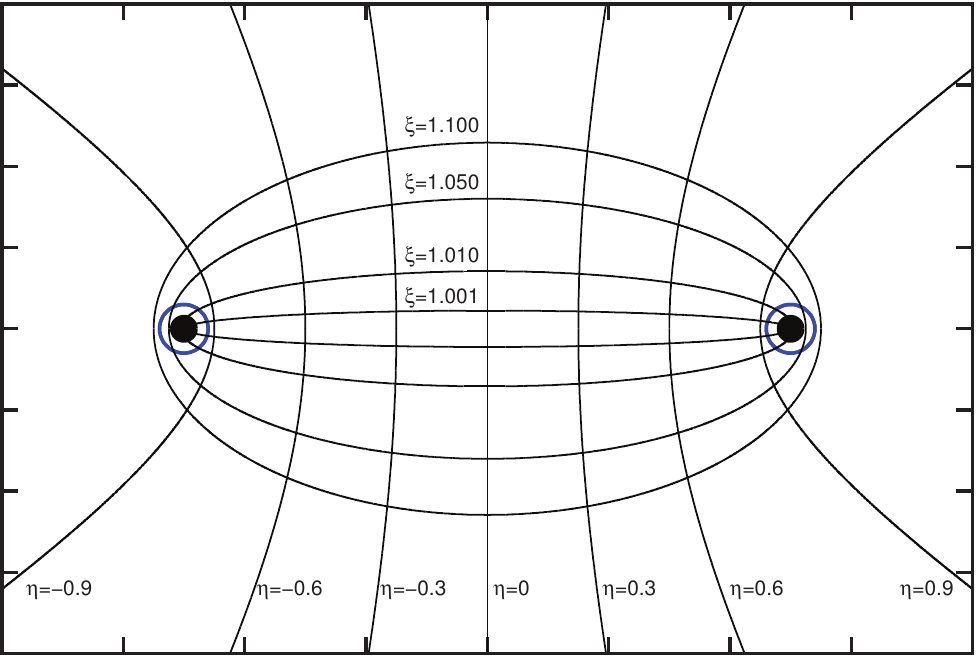}
\caption{The prolate spheroidal coordinate system. The $z$-axis is the internuclear axis of the molecule; $\xi=1$ corresponds to the $z$-axis. Surfaces of constant $\xi$ are ellipsoids, and surfaces of constant $\eta$ are hyperboloids. For $\xi=1$, $\eta=\pm 1$ gives the positions of the two nuclei. When $\eta=0$, a point lies in the midplane passing through the center of the molecule.  }
\label{fig:Sphcoord}
\end{figure}

Accurate quantum mechanical results can be obtained for H$_2^+$ because, within the Born-Oppenheimer approximation, the two-center Coulomb problem is separable in prolate spheroidal coordinates. In this two-center coordinate system, points are defined as intersections of ellipsoids and hyperboloids, as shown in Fig.~\ref{fig:Sphcoord}. The nuclei are positioned on the $z$-axis, and the coordinates depend on the internuclear separation $R$ and the distances $r_1$ and $r_2$ of the electron from each nucleus, such that:
\begin{eqnarray}
\xi&=&\frac{r_1+r_2}{R}\hspace{5mm}\xi\geq1\\
\eta&=&\frac{r_1-r_2}{R} \hspace{5mm}-1\leq\eta\leq 1\\
\phi&=&\tan^{-1}\left({y}/{x}\right).
\end{eqnarray}
Since surfaces of constant $\xi$ are ellipsoids and surfaces of constant $\eta$ are hyperboloids, $\xi$ and $\eta$ are often referred to as ``quasiradial'' and ``quasiangular'' coordinates.

The stationary state wave function $\Psi$ for H$_2^+$ can be expressed as
\begin{equation}
\Psi(\xi,\eta,\phi)=\frac{N_{\epsilon\lambda}}{}\frac{X(\xi)}{\sqrt{\xi^2-1}}\frac{{N}(\eta)}{\sqrt{1-\eta^2}}\frac{e^{im\phi}}{\sqrt{2\pi}},
\end{equation}
where $N_{\epsilon\lambda}$ is a normalization constant and ${X}(\xi)$ and ${N}(\eta)$ satisfy 
\begin{eqnarray}
\left[\frac{d^2}{d\xi^2}+\left\{\gamma^2+\frac{2R\xi\!-\!\lambda}{\xi^2\!-\!1}+\frac{(1\!-\!m^2)}{(\xi^2\!-\!1)^2}\right\}\right]{X}(\xi)=0&& \label{eq:DExi} \\
\left[\frac{d^2}{d\eta^2}+\left\{\gamma^2+\frac{\lambda}{1\!-\!\eta^2}+\frac{(1\!-\!m^2)}{(1\!-\!\eta^2)^2} \right\}\right]{N}(\eta)=0&&
\label{eq:DEeta}
\end{eqnarray}
In Eqs.~\ref{eq:DExi} and \ref{eq:DEeta}, $\gamma^2=\epsilon R^2/2$ depends on the electronic energy $\epsilon$, and $-\left(\gamma^2+\lambda\right)$ is a shared separation constant. Each state is characterized by a set of node numbers $\left\{n_{\eta}, n_{\xi},n_{\phi}\right\}$, with $n_{\eta}$ even for gerade and odd for ungerade states.

The physical significance of $\lambda$ has been elucidated by Erikson and Hill \cite{CoMHill} and Coulson and Joseph \cite{CoMCoulson}. As the internuclear separation $R$ decreases to $0$ and the system becomes He$^+$, $\lambda\rightarrow l(l+1)$. As $R$ approaches infinity, $\lambda$ becomes a component of the eccentricity, or Runge-Lenz, vector. 

Erikson and Hill \cite{CoMHill} also show that the separation constant, $-\left(\gamma^2+\lambda\right)$, is the third constant of the motion; note that within the WKB approximation $-\left(\gamma^2+\lambda-m^2\right)$ is the barrier height in the $\eta$ coordinate. Owing to this extra symmetry, two potential curves that are otherwise identical (e.g. two $^2\Sigma$ states) may cross if they don't have the same $\lambda$ \cite{CoMCoulson}. One example is the crossing of the 2s$\sigma_g$ and 3d$\sigma_g$ states near $R=4~a_0$. Avoided crossings are still possible because the electron moves in a potential that transitions from a single to a double well as $R$ increases \cite{Rost,FanoDiff}. The crossings are exhibited between gerade potential energy curves of constant $n_{\xi}$ and lead up to the three-body breakup limit; {avoided crossings for selected high-lying states are shown in Fig.~\ref{fig:avoidedcross}.}
\ifthenelse{\boolean{twocolumn}}{
\begin{figure}
\subfigure[]{\includegraphics[width=0.9\columnwidth]{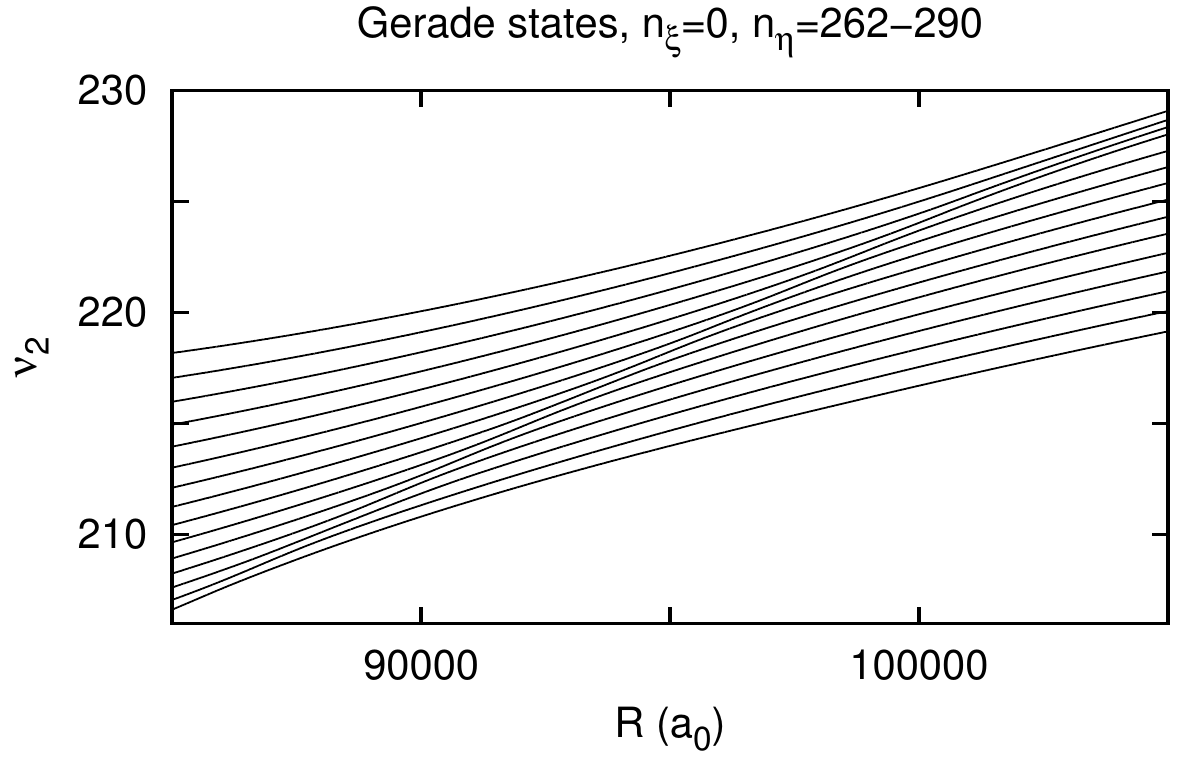}}

\subfigure[]{\includegraphics[width=0.9\columnwidth]{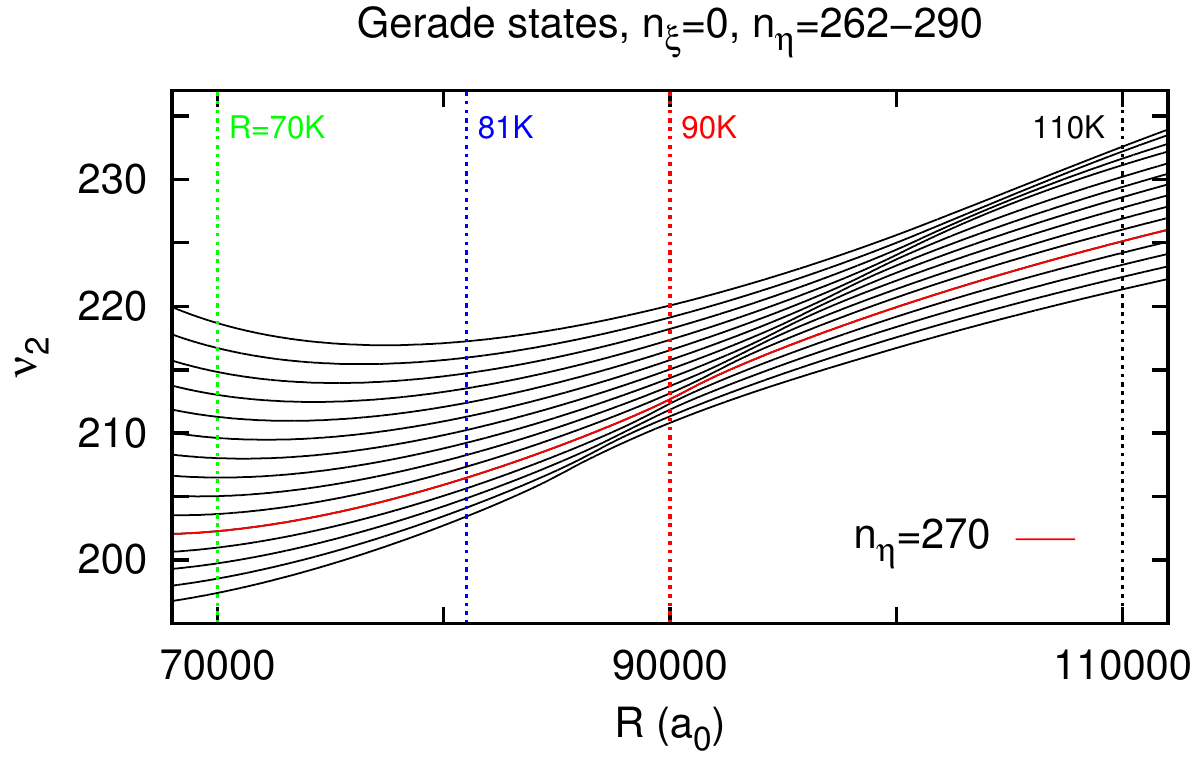}}

\subfigure[]{\includegraphics[width=0.9\columnwidth]{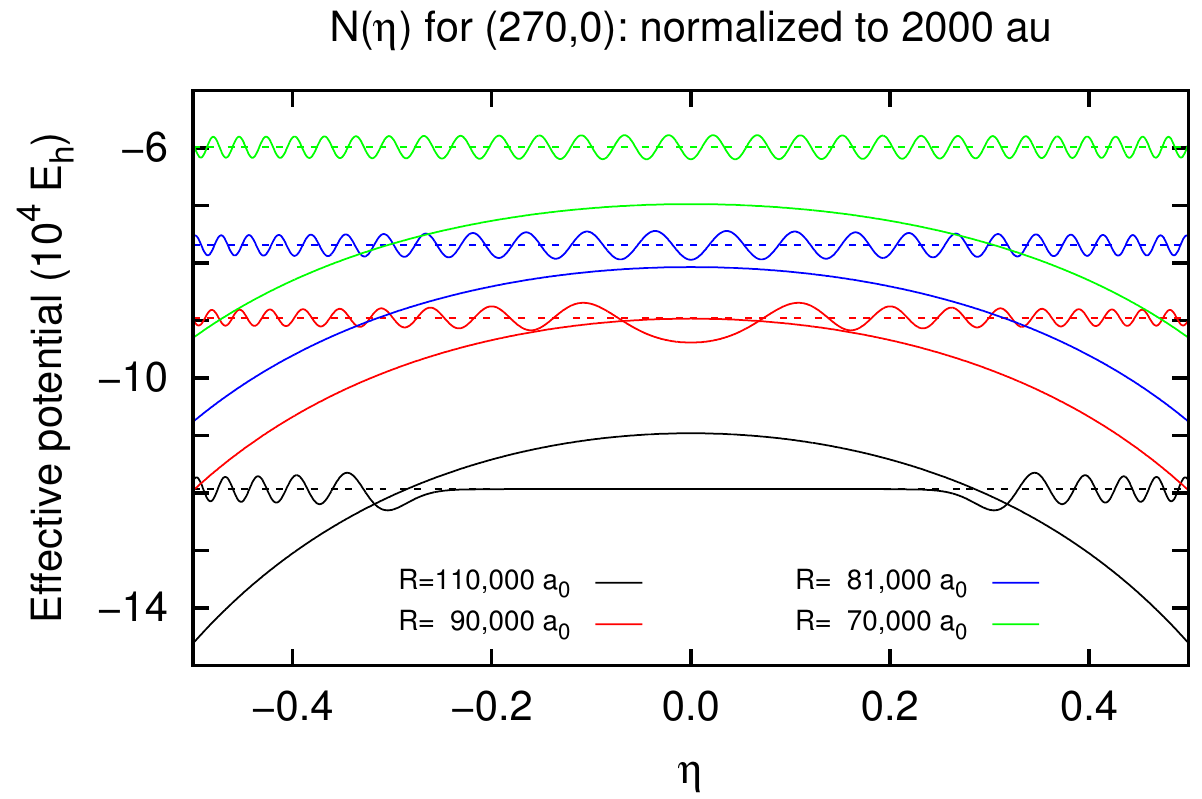}}

\caption{Ridge of avoided crossings shown for states with $n_{\xi}=0$ and $n_{\eta}=262$--290. Panel~(a) shows the $\nu_2(=\sqrt{-2/\epsilon})$ values as a function of $R$; one can clearly see the avoided crossings. Panel (b) highlights in red the curve for the state $(n_{\eta},n_{\xi})=(270,0)$. Panel~(c) shows wave functions for $(n_{\eta},n_{\xi})=(270,0)$ for values of $R$ corresponding to the vertical lines in Panel~(b). As $R$ increases, the potential in which the electron moves transitions from a single to a double well, and the wave function markedly changes character.}
\label{fig:avoidedcross}
\end{figure}}{
\begin{figure}
\subfigure[]{\includegraphics[width=440pt]{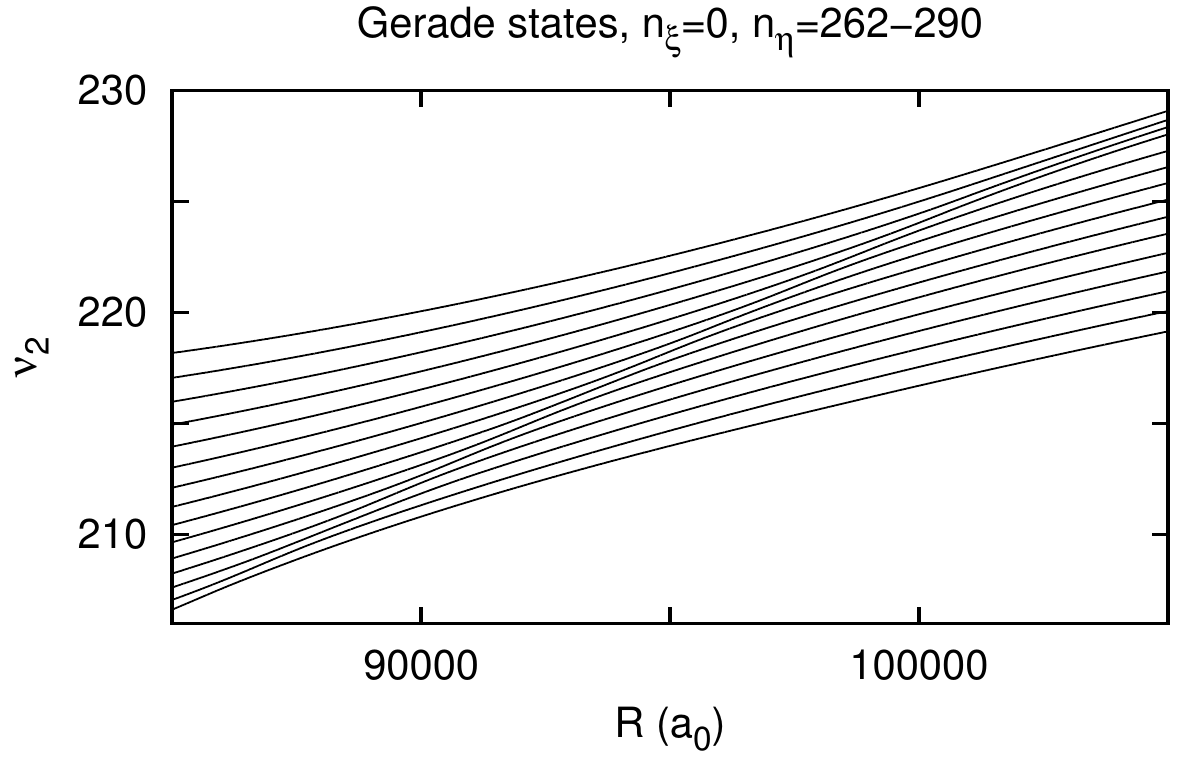}}

\subfigure[]{\includegraphics[width=220pt]{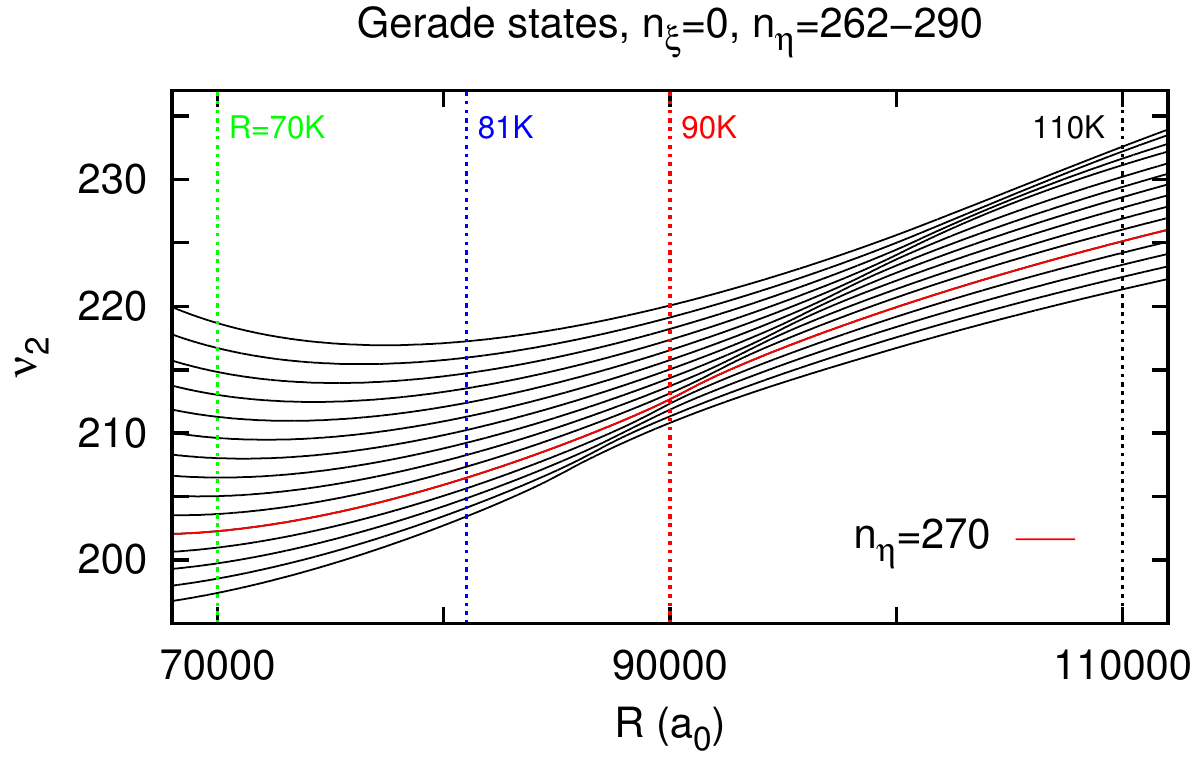}}
\subfigure[]{\includegraphics[width=220pt]{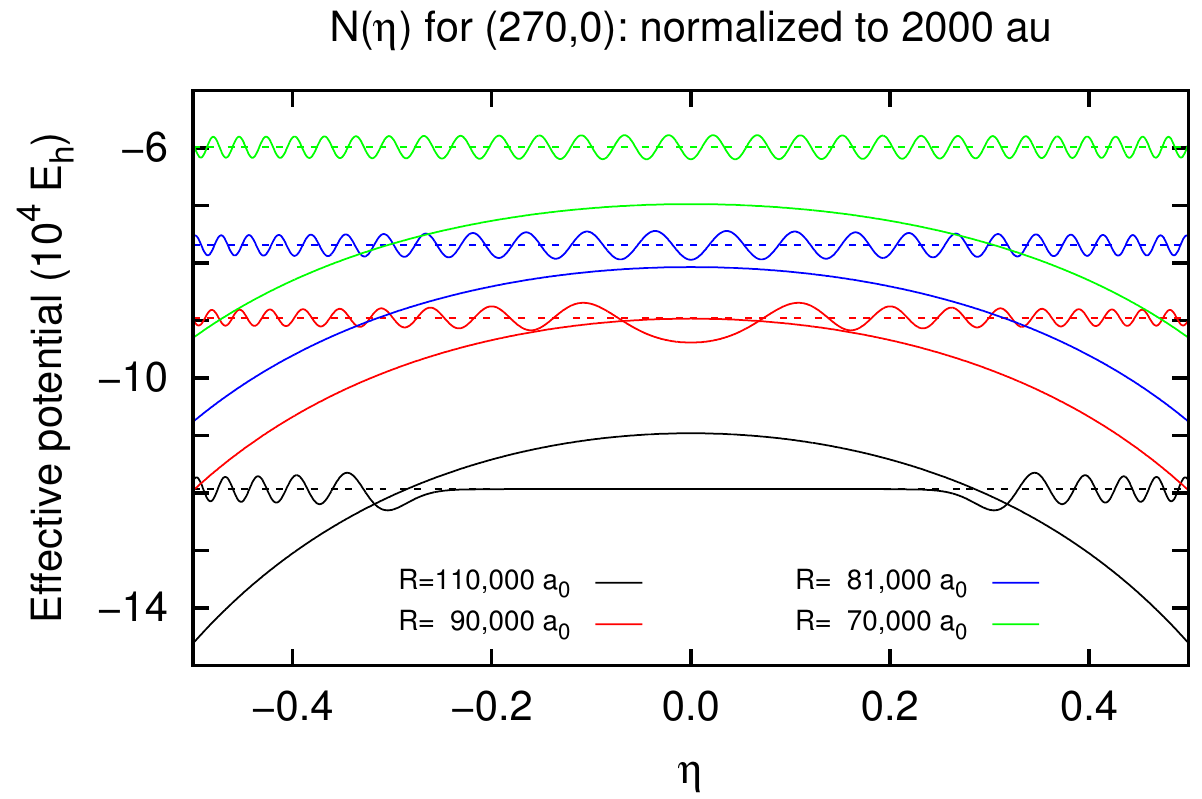}}
\caption{Ridge of avoided crossings shown for states with $n_{\xi}=0$ and $n_{\eta}=262$--290. Panel~(a) shows the $\nu_2(=\sqrt{-2/\epsilon})$ values as a function of $R$; one can clearly see the avoided crossings. Panel~(b) highlights in red the curve for the state $(n_{\eta},n_{\xi})=(270,0)$. Panel~(c) shows wave functions for $(n_{\eta},n_{\xi})=(270,0)$ for values of $R$ corresponding to the vertical lines in Panel~(b). As $R$ increases, the potential in which the electron moves transitions from a single to a double well, and the wave function markedly changes character.}
\label{fig:avoidedcross}
\end{figure}
}

\subsubsection{Numerical details, Estimated Accuracy}
Quantum mechanical calculations for states with $n_{\eta}$ ranging from 0 to 292 have been carried out for states that correspond asymptotically to H $(n\leq 145)$ plus a proton. Our highest energy calculations extend out to  $R\leq 120,000$ $a_0$. 

In these calculations Eq.~\ref{eq:DEeta} is solved for the separation constants of states satisfying either gerade or ungerade boundary conditions at $\eta=0$. This gave a set of possible pairs $(\gamma^2$, $\lambda)$ for each $n_{\eta}$. Next, using those separation constants, Eq.~\ref{eq:DExi} is solved for the energies of states at fixed $R$. The shooting method accurately solves both Eqs.~\ref{eq:DExi} and \ref{eq:DEeta}, with series expansions for $N(\eta)$ near $\eta=-1$ \cite{Baber} and for $X(\xi)$ near $\xi=1$ \cite{Jaffe} matched to numerically propagated wave functions determined by a 4-point predictor corrector method. In solving Eq.~\ref{eq:DExi}, the large $\xi$ boundary condition is imposed where the wave function has decreased by 20 orders of magnitude within a WKB approximation. The numerical integration was performed on a square root mesh, with $x_{\eta}=\sqrt{\eta+1}$ and $x_{\xi}=\sqrt{\xi-1}$. 

Accuracy was assessed by comparing calculated energies with results in the literature for lower-lying states, and by comparing our tabulated $\lambda$ values with those obtained using Mathematica's\cite{Mathematica} algorithm. Madsen and Peek \cite{Madsen} report energies for the 20 lowest states of H$_2^+$ up to $R=150$ $a_0$, and our results typically agree to the 12 decimal places reported, except for states that are high up on the repulsive walls, where the agreement diminishes to 8 digits. Our $\lambda$ values agreed with those given by Mathematica's\cite{Mathematica} algorithm to 8--10 decimal places, except for very large $\lambda$ where that algorithm frequently converges to erroneous values of $\lambda$. 

In addition, for comparisons between phases of the quantum and semiclassical wave functions, we have solved Eqs.~\ref{eq:DExi} and \ref{eq:DEeta} using the Milne phase-amplitude procedure \cite{Milne} with the proper boundary conditions for H$_2^+$ \cite{Larsson}. The Milne phase-amplitude procedure involves solving the differential equation
\begin{equation}
\left[\frac{d^2}{dx^2}+k^2(x)\right]\psi(x)=0 \label{eq:diffeqq}
\end{equation}
exactly by calculating an amplitude $\alpha(x)$ such that the wave function satisfies
\begin{equation}
\psi(x)=N\alpha(x)\sin\left(\int^x \alpha^{-2}(x')\,dx' + \phi_0\right), \label{eq:milnewf}
\end{equation}
where $N$ is a normalization constant, and where $\alpha(x)$ is any particular solution of the non-linear second-order equation
\begin{equation}
\left[\frac{d^2}{dx^2}+k^2(x)\right]\alpha(x)=\alpha^{-3}(x) \label{eq:amplitude}.
\end{equation}
The WKB approximation amounts to replacing $\alpha(x')$ with $1/\sqrt{|k(x')|}$. Equation~\ref{eq:amplitude} can be reexpressed as a third order, linear differential equation \cite{CoteMilne,Kiyokawa}. 

The amplitude $\alpha(x)$ depends on the regular, $\psi(x)$, and an irregular, $\overline{\psi}(x)$, solution of Eq.~\ref{eq:diffeqq} in the following way:
\begin{equation}
\alpha^2(x)=A\left(\psi(x)\right)^2+B\left(\overline{\psi}(x)\right)^2+2C\psi(x)\overline{\psi}(x)\label{eq:expforalph},
\end{equation}
where $\left(AB-C^2\right)^{-1/2}$ is equal to the Wronskian, $W$, between $\psi(x)$ and $\overline{\psi}(x)$. For very small $\xi$ or $\eta$, the regular solutions $X(\xi)$ and $N(\eta)$ approach
\begin{eqnarray}
X(\xi)&\underset{\xi\rightarrow 1}{\rightarrow}&\left(\xi^2\!-\!1\right)^{1/2}\times\mathrm{const.} \label{eq:regxi}\\
N(\eta)&\underset{\eta\rightarrow -1}{\rightarrow}&\left(1\!-\!\eta^2\right)^{1/2}\times\mathrm{const.} \label{eq:regeta}
\end{eqnarray}
and so an irregular solution has the behavior
\begin{eqnarray}
\overline{X}(\xi)\underset{\xi\rightarrow 1}{\rightarrow}\left(\xi^2\!-\!1\right)^{1/2}\ln(\xi^2\!-\!1)\times\mathrm{const.}&& \label{eq:irregxi}\\
\overline{N}(\eta)\underset{\eta\rightarrow -1}{\rightarrow}\left(1\!-\!\eta^2\right)^{1/2}\ln(1\!-\!\eta^2)\times\mathrm{const.}&& \label{eq:irregeta}.
\end{eqnarray}.

The wave function in Eq.~\ref{eq:milnewf} must satisfy the appropriate boundary conditions \cite{Larsson} for H$_2^+$. The first boundary condition, that $X(\xi)$ or $N(\eta)$ are zero when $\xi=1$ or $\eta= -1$, is satisfied by expressing the wave functions as  
\begin{eqnarray}
X(\xi)&=&N_{\xi}\alpha(\xi)\sin\left(\int_{1}^\xi \alpha^{-2}(\xi')\,d\xi'\right)\\
N(\eta)&=&N_{\eta}\alpha(\eta)\sin\left(\int_{-1}^\eta \alpha^{-2}(\eta')\,d\eta'\right) \label{eq:MilneEtawf}
\end{eqnarray}
where $\eta\leq 0$ in Eq.~\ref{eq:MilneEtawf}. The second boundary condition puts a constraint on the total phase and yields quantization conditions in $\eta$ and $\xi$; the total phase is the phase parameter $\beta$ used in quantum-defect theory. The phase of $X(\xi)$ must satisfy the condition
\begin{equation}
\int_1^{\infty} \alpha^{-2}(\xi)\,d\xi=(n_{\xi}+1)\pi \label{eq:quantxi}
\end{equation} 
Above the barrier, the total phase of $N(\eta)$ satisfies
\begin{equation}
2\int_{-1}^0 \alpha^{-2}(\eta)\,d\eta=(n_{\eta}+1)\pi \label{eq:quanteta1}
\end{equation}
while below the barrier
\ifthenelse{\boolean{twocolumn}}{
\begin{eqnarray}
2\int_{-1}^0 \alpha^{-2}(\eta)\,d\eta=(n_{\eta}+1)\pi&& \text{or} \label{eq:quanteta2}\\
\int_{-1}^0 \alpha^{-2}(\eta)\,d\eta=\left(\frac{n_{\eta}}{2}+1\right)\pi&&\label{eq:quanteta3}\\-\tan^{-1}\!\left(\frac{2\alpha(0)}{\left.d\alpha^{-2}/d\eta\right|_{\eta=0}}\right)&& \nonumber
\end{eqnarray}}{
\begin{eqnarray}
2\int_{-1}^0 \alpha^{-2}(\eta)\,d\eta&=&(n_{\eta}+1)\pi \hspace{5mm} \text{or} \label{eq:quanteta2}\\
\int_{-1}^0 \alpha^{-2}(\eta)\,d\eta&=&\left(\frac{n_{\eta}}{2}+1\right)\pi\!-\!\tan^{-1}\!\left(\frac{2\alpha(0)}{\left.d\alpha^{-2}/d\eta\right|_{\eta=0}}\right) \label{eq:quanteta3}
\end{eqnarray}}
depending on whether the state is ungerade (Eq.~\ref{eq:quanteta2}) or gerade (Eq.~\ref{eq:quanteta3}).

To solve for the Milne function, the third order, linear differential equation \cite{CoteMilne,Kiyokawa} is solved for the amplitude $\alpha$ from a point in the classical region \cite{YooandGreene} with a 4-point predictor corrector method; for the boundary condition at this point we used {an iterative method introduced} by Seaton and Peach \cite{SandP}. The correct phase of the solution for small $\xi$ or $\eta$ near the singular points is ensured by implementing the variable transformations $x_{\xi}=\log(\xi-1)$ and $x_{\eta}=\log(\eta+1)$, integrated from $x_0=-350$; then a formula of Korsch and Laurent \cite{Korsch} evaluates the contribution to the phase from $x<-350$. Korsch and Laurent's formula \cite{Korsch} is given by
\begin{equation}
\int \alpha^{-2}(x)\,dx=-\tan^{-1}\left(WC+\frac{WA\psi(x)}{\overline{\psi}(x)}\right)+\mathrm{const.} \label{eq:Korsch}
\end{equation}
where $A$ and $C$ are constants in Eq.~\ref{eq:expforalph}. After substituting Eqs.~\ref{eq:regxi} and \ref{eq:irregxi} or Eqs.~\ref{eq:regeta} and \ref{eq:irregeta} into Eq.~\ref{eq:expforalph}, one finds that the phase integral
\begin{eqnarray}
\int \alpha^{-2}(\xi)\,d\xi&=&\mathrm{const}\times\int \frac{d\xi}{(\xi^2-1)\left(\ln(\xi^2-1)+\mathrm{const}\right)^2} \hspace{4mm} \text{or}\\
\int \alpha^{-2}(\eta)\,d\eta&=&\mathrm{const}\times\int \frac{d\eta}{(1-\eta^2)\left(\ln(1-\eta^2)+\mathrm{const}\right)^2}
\end{eqnarray}
has the form 
\begin{equation}
\int_{-\infty}^{x_0} \frac{dx}{ax^2+bx+c}=\left(2\tan^{-1}\left(\frac{2ax_0+b}{\sqrt{4ac-b^2}}\right)+\pi\right)/\sqrt{4ac-b^2} \label{eq:myint}.
\end{equation}
where $x$ is either $x_{\xi}$ or $x_{\eta}$, and where the right-hand side of Eq.~\ref{eq:myint} has the form of Eq.~\ref{eq:Korsch}. The constants $a$, $b$, and $c$ are readily determined from a least-squares fit\cite{LMdriver} to $\exp(x)\alpha^{-2}(x)$. 

Our main purpose here is to compare to semiclassical results, so the calculations presented here are based on the accurate separation constants already obtained with the shooting method. (One could also determine these constants by using the {Milne} quantization conditions in $\eta$, Eqs.~\ref{eq:quanteta1}--\ref{eq:quanteta3}.) %With the tabulated separation constants, we optimized the point at which the boundary condition in the classical region was imposed so that the quantization conditions in $\xi$ and $\eta$ {were satisfied to} 9 or 10 decimal places, except for the $n_{\xi}=0$ state. 

\subsection{Semiclassical WKB Treatment}
The semiclassical WKB treatment of H$_2^+$ is relatively straightforward, but there are two complications that have been discussed in the literature. The first is that the ranges of $\xi$ and $\eta$ do not extend from $-\infty$ to $\infty$; this is overcome by making a Langer-type modification to the terms in the curly brackets of Eqs.~\ref{eq:DExi} and \ref{eq:DEeta}, which we refer to as the ``local momenta.'' The second difficulty arises for $^2\Sigma$ states, because even the modified local momenta, or ``quasimomenta,'' exhibit simple poles in the classically accessible regions. The present section discusses the derivation of quantization conditions for $^2\Sigma$ states of H$_2^+$ in light of these complications.  

As was pointed out by Strand and Reinhardt \cite{Reinhardt} and by Gershtein \latin{et al.}\ \cite{Gershtein}, Langer-type corrections to the local momenta in Eqs.~\ref{eq:DExi} and \ref{eq:DEeta} are not unique. For instance, the transformations $x=\tan\left(\pi\eta/2\right)$ and $x=\tanh^{-1}(\eta)$ both involve a range of $x$ that extends from $-\infty$ to $\infty$, but each leads to a different quasimomentum. To avoid this issue, Strand and Reinhardt perform an analysis using canonical invariants \cite{Reinhardt}. On the other hand, Gershtein \latin{et al.}\ use quasimomenta which agree with the momenta of the classical Hamilton-Jacobi (HJ) equation:
\begin{eqnarray}
k_{\mathrm{L}}^2({\eta})&=&\frac{-m^2}{(1-\eta^2)^2}+\gamma^2+\frac{\lambda}{1-\eta^2} \label{eq:kleta}\\
k_{\mathrm{L}}^2({\xi})&=&\frac{-m^2}{(\xi^2-1)^2}+\gamma^2+\frac{2R\xi-\lambda}{\xi^2-1},
\label{eq:klxi}
\end{eqnarray}
The quasimomenta in Eqs.~\ref{eq:kleta} and \ref{eq:klxi} are sometimes referred to as ``Bethe-modified,'' and we will use these forms in our analysis. (Strand and Reinhardt \cite{Reinhardt} obtain wave functions in the classical region with the same form as in Gershtein \latin{et al.}\ \cite{Gershtein}.) 

Different procedures for determining Langer-type modifications for general systems have been discussed \cite{Adams,Collins,Gu} and are not our focus here, but we do wish to make a few comments on this connection to the HJ equation. Farrelly and Reinhardt \cite{Farrelly} have shown that for hydrogen in a uniform electric field, the usual quasimomenta in either parabolic and squared parabolic coordinates, in which $-m^2$ replaces $(1-m^2)$, also agree with the momenta in the separated HJ equation. Moreover, the Langer modification for the hydrogen atom in spherical coordinates gives a radial quasimomentum that agrees with the corresponding quantity in the HJ equation, provided one identifies the classical angular momentum $L$ with $(l+1/2)\hbar$ \cite{Berry}. In general, for a separated, time independent 3D problem, using such quasimomenta leads to a total wave function that depends on the classical action in the following way. Suppose the WKB wave function for each separate coordinate $x_k$ is $e^{iS_k(x_k)/\hbar}$, with $p_k^2(x_k)=(dS_k(x_k)/dx_k)^2$ in the zeroth order of $\hbar/i$. If the corrected quasimomenta $p_k(x_k)$ are the same as the momenta in the separated HJ equation, then one can identify the classical action as the sum $S(x_1,x_2,x_3,t)=S_1(x_1)+S_2(x_2)+S_3(x_3)-Et$; the total WKB wave function $e^{iS(x_1,x_2,x_3,t)/\hbar}$ is expressed in terms of the classical action and is partitioned in a very natural way, $e^{i\sum_k S_k(x_k)/\hbar}$. Of course in general the WKB wave function for each coordinate is some linear combination of $e^{\pm iS_k(x_k)/\hbar}$, and there can be additional cross terms that appear in the total wave function.

\begin{figure}[!ht]
\centering
\subfigure[]{\includegraphics[width=3.0in]{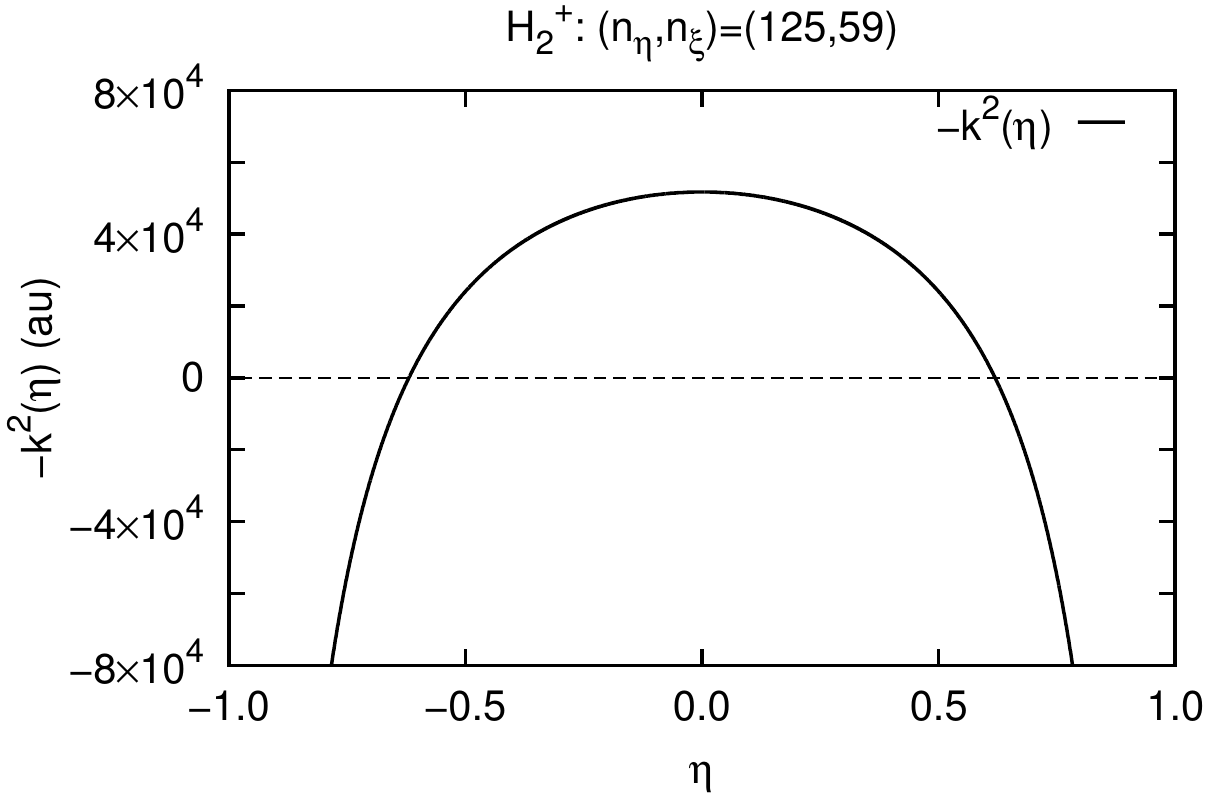}}
\hspace{8mm}\subfigure[]{\includegraphics[width=3.0in]{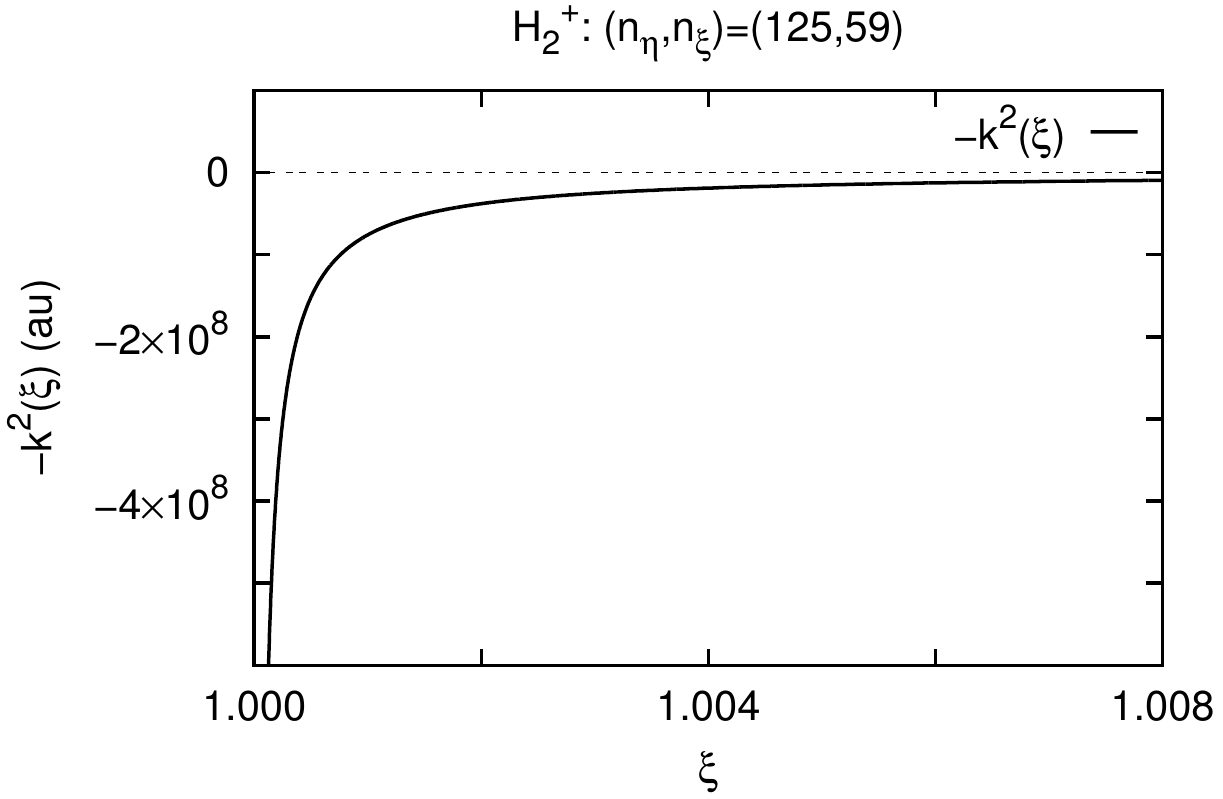}}
\caption{Quasimomenta in $\eta$, Panel~(a), and $\xi$, Panel~(b) for a $^2\Sigma$ state of H$_2^+$. }
\label{fig:potxiandeta}
\end{figure}

We now consider the quasimomenta {in Eqs.~\ref{eq:kleta} and \ref{eq:klxi}}. For $^2\Sigma$ states, there are two roots $\eta_{\mathrm{c}}$ of Eq.~\ref{eq:kleta}:
\begin{equation}
\eta_{\mathrm{c}}=\pm\sqrt{\frac{\gamma^2+\lambda}{\gamma^2}}.
\end{equation}
When $\gamma^2+\lambda$ is negative, the $\eta_{\mathrm{c}}$ correspond to two real roots. Since $\lambda\geq 0$ for H$_2^+$, these roots are always between $\eta=\pm 1$. As $\gamma^2+\lambda$ approaches zero, the two roots coalesce to $\eta_{\mathrm{c}}=0$, and the $\eta_{\mathrm{c}}$ are no longer simple zeros of Eq.~\ref{eq:kleta}. When $\gamma^2+\lambda$ is positive, the roots $\eta_{\mathrm{c}}$ are purely imaginary complex conjugates. In other words, there is a potential barrier in the $\eta$ coordinate with the separation constant, $-(\gamma^2+\lambda)$, as its height; the separation constant is the third constant of the motion for the system. The barrier in the $\eta$ coordinate depends on $R$, $\epsilon$, and the number of nodes in $N(\eta)$. An example in which the electron is below the barrier is shown in Panel~(a) of Fig.~\ref{fig:potxiandeta}. Note that the poles at $\eta=\pm 1$ are in the classically accessible regions.

In contrast, the quasimomentum in $\xi$, defined by Eq.~\ref{eq:klxi}, does not exhibit a potential barrier. Instead, it has either one or no minimum. If $\lambda\leq 2R$, then there is no minimum in $k_{\mathrm{L}}^2(\xi)$, and the pole at $\xi=1$ is in the classical region. If $\lambda>2R$, there is a single minimum which bars the electron from the region near the internuclear axis. This minimum moves further out in $\xi$ as $\lambda$ increases past $2R$. For the case where $R=0$, that is, for He$^+$, $\lambda$ is $l(l+1)$ and this minimum in $\xi$ is associated with the combined centrifugal and Coulomb potentials. In the general case of nonzero $R$, for a fixed energy, $\lambda$ increases with $n_{\eta}$; with sufficient ``angular'' excitation, a minimum appears in $k_{\mathrm{L}}^2(\xi)$. 

One can also show that for there to be a minimum in $k^2_{\mathrm{L}}(\xi)$, that is, for there to be two physical roots of
\begin{equation}
\xi_{\mathrm{c}}^2+\left({2R}/{\gamma^2}\right)\xi_{\mathrm{c}}-\eta_{\mathrm{c}}^2 = 0,
\end{equation}
then the electron must be above the potential barrier. Moreover, for $R>0$, $\nu_2$ must be greater than $\sqrt{R/2}$; for large $R$, this occurs at small electronic energies $|\epsilon|$, typically high on the repulsive wall of the potential. 
More often than not we find there is no minimum in the potential, and an example is shown in Panel~(b) of Fig.~\ref{fig:potxiandeta}.

For $^2\Sigma$ states, then, there are first-order poles in the quasimomenta at $\eta=\pm 1$ and $\xi=1$ which are typically in the classical regions. Near these poles the WKB assumptions break down. Athavan \latin{et al.}\ \cite{Froman1} have rigorously derived the connection formula for this case. Far enough from the pole for the WKB approximation to be valid, the wave function for either $\eta$ or $\xi$, which they denote $\psi$, is given by:
\begin{equation}
\psi(x)\propto \frac{1}{\sqrt{k(x)}}\sin\left(\int^x k(x')\,dx'+\frac{\pi}{4}\right).
\label{eq:Froman}
\end{equation}
Thus the phase accumulation of the wave function equals $\pi/4$ near the pole, and in the classical region, far enough from the pole for the WKB assumptions to be satisfied, the wave function has the same form as if the poles were replaced by classical turning points. 

Given Eq.~\ref{eq:Froman}, the quantization condition in $\xi$ is the same as it would be for a single well potential. The same is true for $\eta$ when the electronic energy is far enough above the potential barrier that the reflection probability is negligible. Therefore we have
\begin{eqnarray}
\int_{\xi_{\mathrm{c}1}}^{\xi_{\mathrm{c}2}} k_{\mathrm{L}}({\xi})\,d\xi&=&\pi(n_{\xi}+1/2) \label{eq:q1} \\
2\int_{-1}^0 k_{\mathrm{L}}({\eta})\,d\eta&=&\pi(n_{\eta}+1/2)
\label{eq:q2}
\end{eqnarray}
where $\xi_{\mathrm{c}2}$ is the position of the outermost classical turning point in $\xi$ and $\xi_{\mathrm{c}1}=1$ when $k^2_{\mathrm{L}}(\xi)$ has no minimum. Gershtein \latin{et al.}\ determine a quantization condition that includes reflection above the barrier in $\eta$. It is more accurate, however, to take the approach of Strand and Reinhardt \cite{Reinhardt} and use the uniform method to account for the fact that the turning points can be close together. For the $\xi$ coordinate, a similar approach could be used near the bottom of the well to improve results there.

Below the potential barrier, the quantization conditions in $\eta$ are the same as for a double well potential. One requires that in the forbidden region  
\begin{eqnarray}
N(\eta)&\propto&\frac{\sinh\left(\int_0^{\eta} |k_{\mathrm{L}}(\eta)| \, d\eta\right)}{\sqrt{|k_{\mathrm{L}}(\eta)|}}\hspace{5mm}\text{or} \label{eq:eta0bc} \\
N(\eta)&\propto&\frac{\cosh\left(\int_0^{\eta} |k_{\mathrm{L}}(\eta)|\,d\eta\right)}{\sqrt{|k_{\mathrm{L}}(\eta)|}} \label{eq:eta0bc2}
\end{eqnarray}
for ungerade or gerade states, respectively. By using the standard connection formulae for isolated classical turning points, along with Eq.~\ref{eq:Froman}, one can show that in order for the wave function in the forbidden region to have the form of Eq.~\ref{eq:eta0bc}, one must require
\begin{equation}
\cot\left(\int_{-1}^{-\eta_{\mathrm{c}}}k_{\mathrm{L}}(\eta)\right)={\pm{\exp\left({-2\int_{-\eta_{\mathrm{c}}}^0|k_{\mathrm{L}}(\eta)|}\right)}/{2}}
\label{eq:Tun}
\end{equation}
where there is a minus sign on the right hand side for ungerade states. This is the same quantization condition as given by Gershtein \latin{et al.}\ \cite{Gershtein}. When the energy is far enough below the barrier, the right-hand side of Eq.~\ref{eq:Tun} is negligible, and the quantization conditions reduce to  
\begin{eqnarray}
2\int_{-1}^{\eta_c}k_{\mathrm{L}}({\eta})\,d\eta&=&\pi (n_{\eta}+1)\hspace{2 mm}\text{or} \label{eq:notun}\\
&=&\pi n_{\eta} \label{eq:notun2}
\end{eqnarray}
depending on whether the state has gerade (Eq.~\ref{eq:notun}) or ungerade (Eq.~\ref{eq:notun2}) symmetry.

In the classical region, the WKB wave functions have the form 
\begin{eqnarray}
X(\xi)&\propto&\sqrt{\frac{1}{k_{\mathrm{L}}(\xi)}}\sin\left(\int_{\xi_{\mathrm{c}1}}^{\xi_{\mathrm{c}2}}k_{\mathrm{L}}(\xi)+\pi/4\right)\\
N(\eta)&\propto&\sqrt{\frac{1}{k_{\mathrm{L}}(\eta)}}\sin\left(\int_{-1}^{-\eta_c}k_{\mathrm{L}}(\eta)+\pi/4\right),
\end{eqnarray}
where the $\eta$ wave function is either symmetric or antisymmetric about $\eta=0$. These forms agree with those given by Strand and Reinhardt \cite{Reinhardt} and Gershtein \latin{et al.}\ \cite{Gershtein}.

Near the classical turning points, when they are not too close together, the quasimomentum $k^2_{\mathrm{L}}$ can be defined to be $2\alpha(x-x_{\mathrm{c}})$ and{, with} $z:=-(2\alpha)^{1/3}(x-x_{\mathrm{c}})$, one has
\begin{eqnarray}
X(\xi)&\propto&\frac{\sqrt{\pi}}{\left(2\alpha\right)^{1/6}}\sin\left({\int_{\xi_{\mathrm{c}1}}^{\xi_{\mathrm{c}2}}k_{\mathrm{L}}(\xi)}\right)\mathrm{Ai}(z)\\
N(\eta)&\propto&\frac{\sqrt{\pi}}{\left(2\alpha\right)^{1/6}}\left\{\sin\left({\int_{-1}^{-\eta_c}k_{\mathrm{L}}(\eta)}\right)\mathrm{Ai}(z)\right.\nonumber\\
&&\hspace{16mm}+\left.\cos\left({\int_{-1}^{-\eta_c}k_{\mathrm{L}}(\eta)}\right)\mathrm{Bi}(z)\right\} \label{eq:AiryEta}
\end{eqnarray}
where Ai$(z)$ and Bi$(z)$ are the Airy functions of the first and second kind, and where Eq.~\ref{eq:AiryEta} applies to states below the potential barrier.

Finally, in the forbidden region
\begin{equation}
X(\xi)\propto\frac{1}{\sqrt{\left|k_{\mathrm{L}}(\xi)\right|}}\exp\left({-\int_{\xi_{\mathrm{c}2}}^\infty \left|k_{\mathrm{L}}(\xi)\right|\,d\xi}\right)
\end{equation}
and $N(\eta)$ has the form of Eq.~\ref{eq:eta0bc} for ungerade or of Eq.~\ref{eq:eta0bc2} for gerade states. 

\section{Results and Discussion}
\subsection{Energies}
Figure~\ref{fig:corrdia} shows a comparison between our semiclassical and quantum mechanical values of $\nu_2(=\sqrt{-2/\epsilon})$ for low-lying states of H$_2^+$. Panel~(a), which shows the comparison for a smaller range than Panel~(b), illustrates where the tunneling rate is large enough that the right-hand side of Eq.~\ref{eq:Tun} is non-negligible; results obtained by using the full Eq.~\ref{eq:Tun} are labeled ``WKB,'' while results {obtained by using Eqs.~\ref{eq:notun} or \ref{eq:notun2}} are labeled ``WKB2.'' Of course, it's only important to account for the true tunneling rate by using Eq.~\ref{eq:Tun} near the barrier tops, where the gerade-ungerade splitting is just starting to manifest itself in the figure. There are bumps, however, in the ``WKB'' energies very close to the barrier tops where the turning points cannot be considered isolated. In Panel (a) of Fig.~\ref{fig:corrdia} one can see, for instance, a bump near $R=3.5~a_0$ for the state ($n_{\eta},n_{\xi})=(1,0)$ where the $\nu_2$ value is overestimated. Accounting for reflections just above the barrier, as was done by Gershtein \latin{et al.}\ \cite{Gershtein}, diminishes but does not remove the bumps near the barrier tops, {which can be seen in their Fig.~4.}
\ifthenelse{\boolean{twocolumn}}{
\begin{figure}[!ht]
\subfigure[]{\includegraphics[width=0.95\columnwidth]{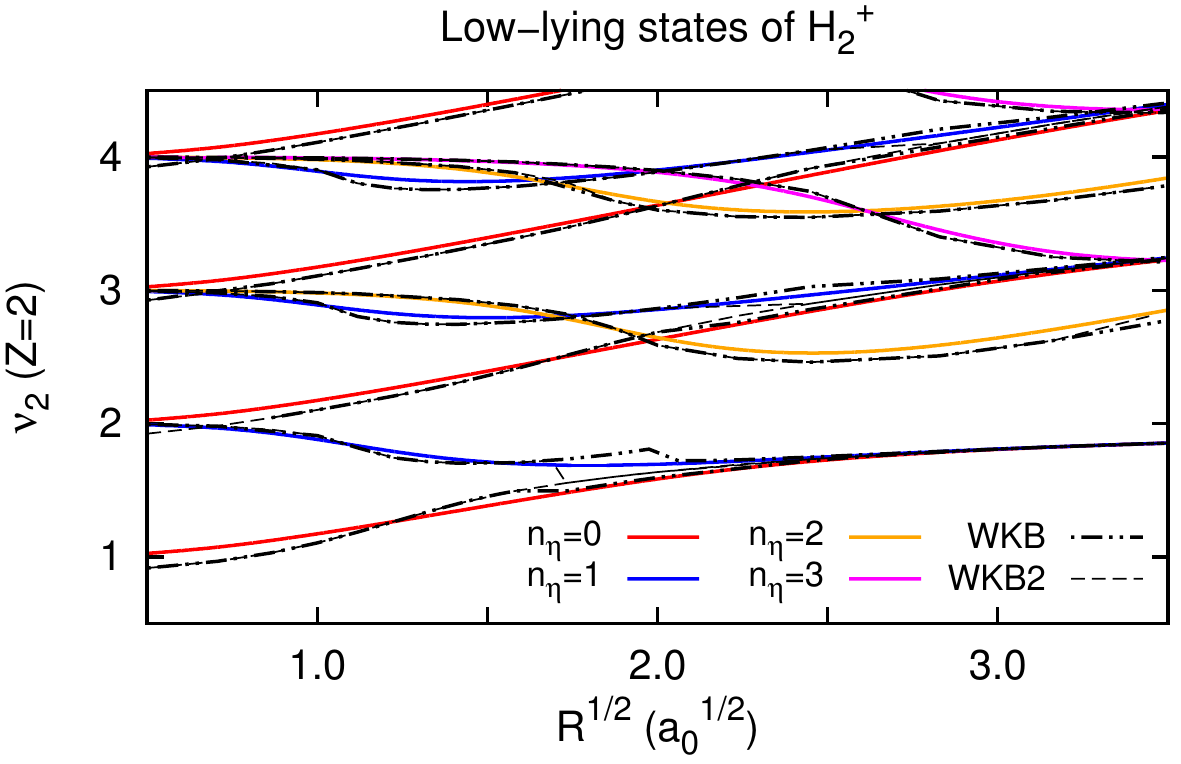}}

\subfigure[]{\includegraphics[width=0.95\columnwidth]{LowStates-QMvsWKB-Expand.pdf}}
\caption{Comparison between quantum (solid lines) and semiclassical (dashed and center lines) $\nu_2$ values for $^2
\Sigma$ states of H$_2^+$. Panel~(a) shows the comparison for low-lying states at relatively small $R$. The label ``WKB2'' refers to the method by neglecting tunneling in Eq.~\ref{eq:Tun}, while ``WKB'' refers to results obtained with the full Eq.~\ref{eq:Tun}. Panel~(b) shows the comparison with the ``WKB'' results for a wider range of $R$ and $\nu_2$.} 
\label{fig:corrdia}
\end{figure}}
{
\begin{figure}[!ht]
\subfigure[]{\includegraphics[width=4.5in]{LowStates-QMvsWKB.pdf}}

\subfigure[]{\includegraphics[width=4.5in]{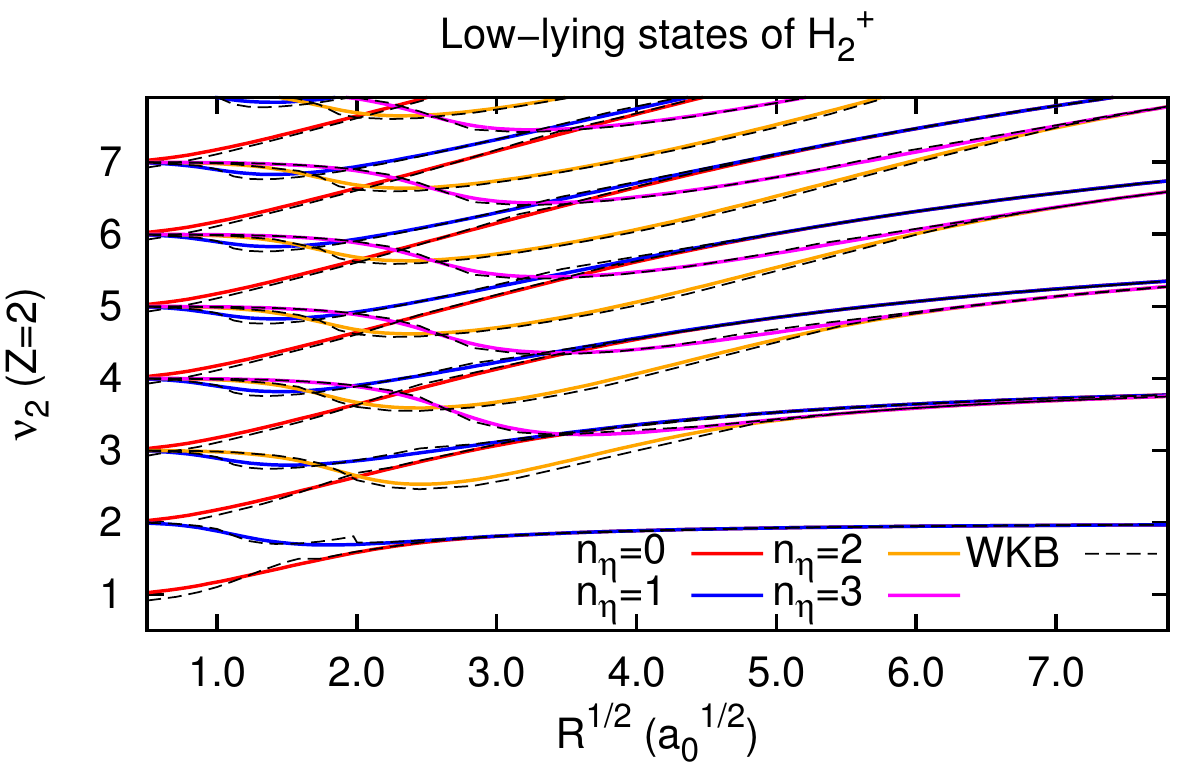}}
\caption{Comparison between quantum (solid lines) and semiclassical (dashed and center lines) $\nu_2$ values for $^2
\Sigma$ states of H$_2^+$. Panel~(a) shows the comparison for low-lying states at relatively small $R$. The label ``WKB2'' refers to the method by neglecting tunneling in Eq.~\ref{eq:Tun}, while ``WKB'' refers to results obtained with the full Eq.~\ref{eq:Tun}. Panel~(b) shows the comparison with the ``WKB'' results for a wider range of $R$ and $\nu_2$.} 
\label{fig:corrdia}
\end{figure}
}

Strand and Reinhardt \cite{Reinhardt} account for the proximity of the turning points near the height of the barrier with a uniform approach. Panel~(a) of Fig.~\ref{fig:CompRandS} compares the $\nu_2$ values they obtained by using the uniform approach (dashed lines) with our WKB results (dotted lines), WKB results \cite{Reinhardt} that neglect tunneling in Eq.~\ref{eq:Tun} (solid circles), and the quantum values (solid lines). Accurately determining the tunneling integral in Eq.~\ref{eq:Tun} improves results below the barrier top. Then the energies can be further refined by using a uniform approach, which gives better results where the turning points are close together. Panel~(b) of Fig.~\ref{fig:CompRandS} shows the difference between the quantum mechanical and semiclassical values of $\nu_2$ for the $1$p$\sigma_u$ state. Energies above the potential barrier lie to the left of the vertical line, while energies below the barrier lie to the right. Near $\sqrt{R}=1.8$, the energy is very close to the height of the barrier, and the uniform approach gives better agreement just above and below the top.
\ifthenelse{\boolean{twocolumn}}{
\begin{figure}[!ht]
\centering
\subfigure[]{\includegraphics[width=0.95\columnwidth]{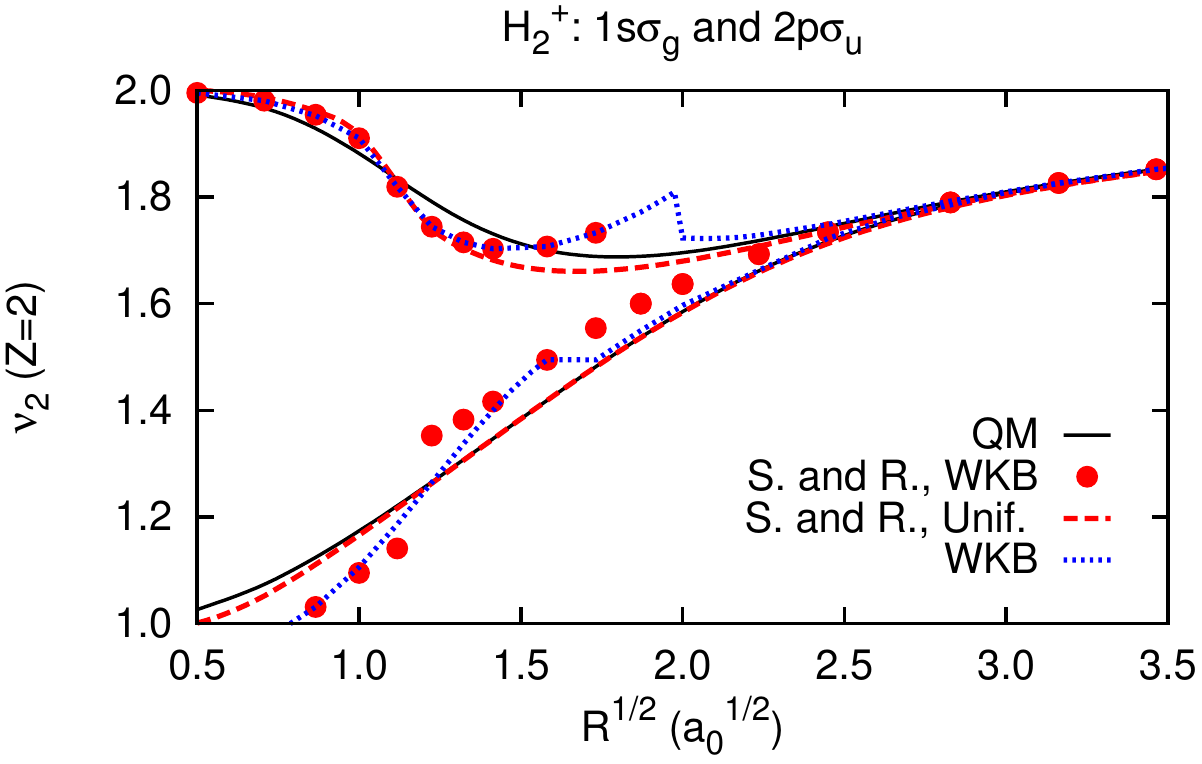}}

\subfigure[]{\includegraphics[width=0.95\columnwidth]{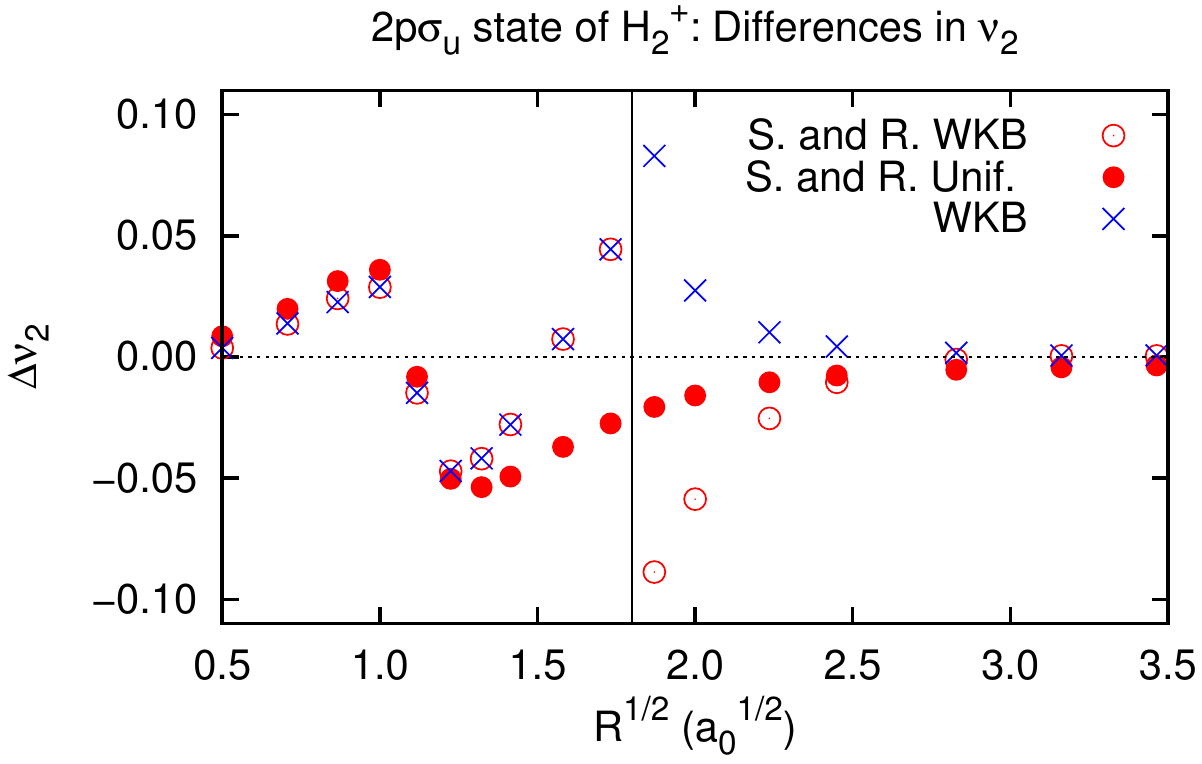}}
\caption{Comparison between quantum (``QM'') and semiclassical $\nu_2$ calculated by Strand and Reinhardt \cite{Reinhardt} (``S. and R.'') and ourselves (``WKB''). Panel~(a) shows the $\nu_2$ values versus $\sqrt{R}$, and Panel~(b) shows the deviation of the semiclassical $\nu_2$ from the quantum values for the 1p$\sigma_u$ state of H$_2^+$.}
\label{fig:CompRandS}
\end{figure}}
{
\begin{figure}[!ht]
\centering
\subfigure[]{\includegraphics[width=4.5in]{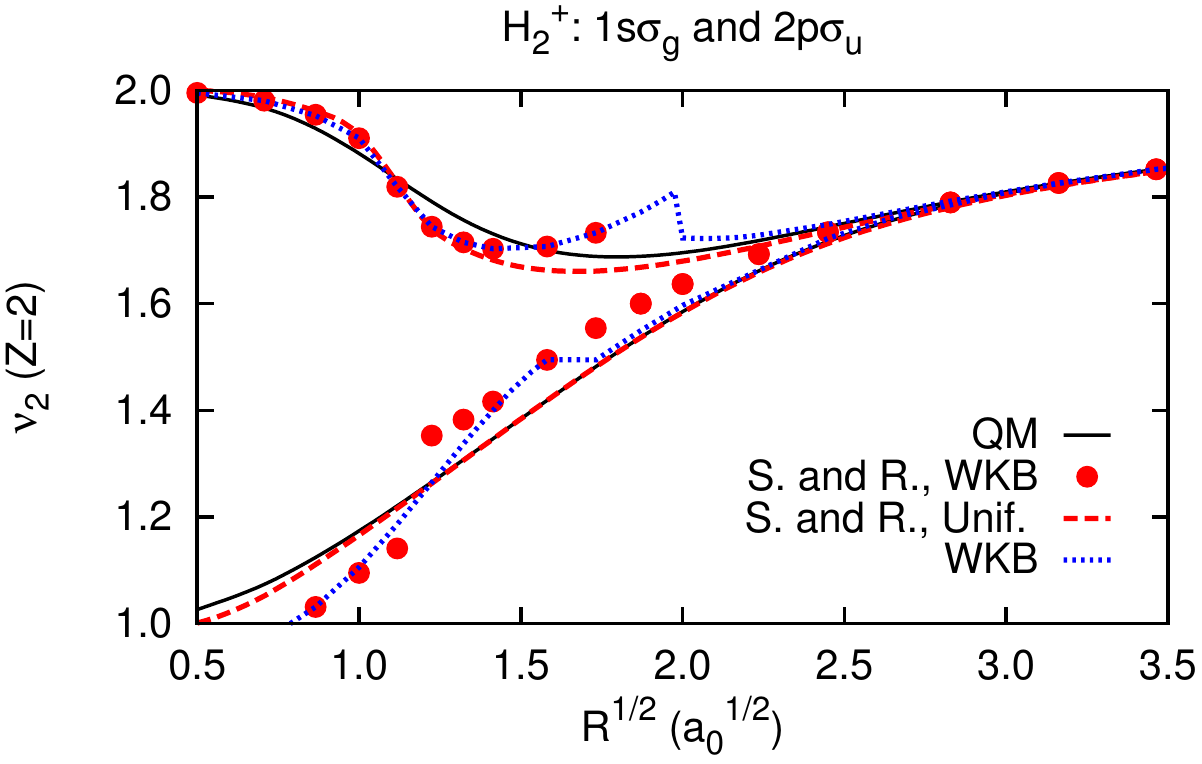}}

\subfigure[]{\includegraphics[width=4.5in]{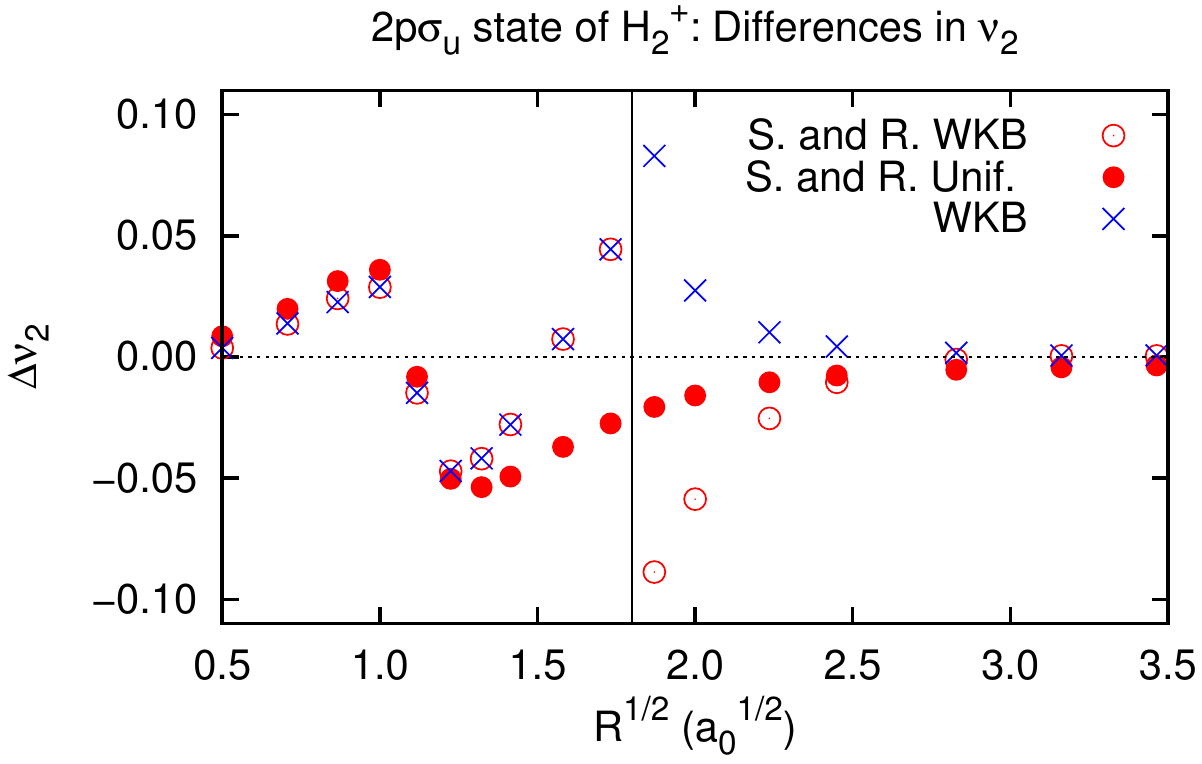}}
\caption{Comparison between quantum (``QM'') and semiclassical $\nu_2$ calculated by Strand and Reinhardt \cite{Reinhardt} (``S. and R.'') and ourselves (``WKB''). Panel~(a) shows the $\nu_2$ values versus $\sqrt{R}$, and Panel~(b) shows the deviation of the semiclassical $\nu_2$ from the quantum values for the 1p$\sigma_u$ state of H$_2^+$.}
\label{fig:CompRandS}
\end{figure}
}

These low-lying states present a worst-case scenario for the WKB approximation, which is why such refined methods are necessary there. As the energies increase, the Born-Oppenheimer potential curves span a wider range of internuclear separations. For instance, the state $(n_{\eta},n_{\xi})=(233,17)$ exhibits a potential minimum near $R=62700$~$a_0$, where $\nu_2\approx 198.5$. For this state, the barrier top is around $R=76300$~$a_0$ and only a small part of the potential curve is affected by it. Panel~(a) of Fig.~\ref{fig:comparepec2} shows the semiclassical and quantum mechanical PECs for this state and five others. Generally the $\nu_2$ values agree to three or four decimal places, as shown in Panel~(b) of Fig.~\ref{fig:comparepec2} for the state $(n_{\eta},n_{\xi})=(233,17)$. This allows one to determine spectroscopic constants for the potentials with high accuracy. In Table~\ref{tab:spectconst}, we compare our estimated spectroscopic constants and find they differ by less than 0.3\%. Finally, Panel~(c) shows that the error in energy caused by the barrier top is localized to a relatively small region of $R$; the position of the barrier top is marked by a vertical line.
\ifthenelse{\boolean{twocolumn}}{
\begin{figure}[!ht]
\subfigure[]{\includegraphics[width=0.95\columnwidth]{PECs-highstates.pdf}}

\subfigure[]{\includegraphics[width=0.95\columnwidth]{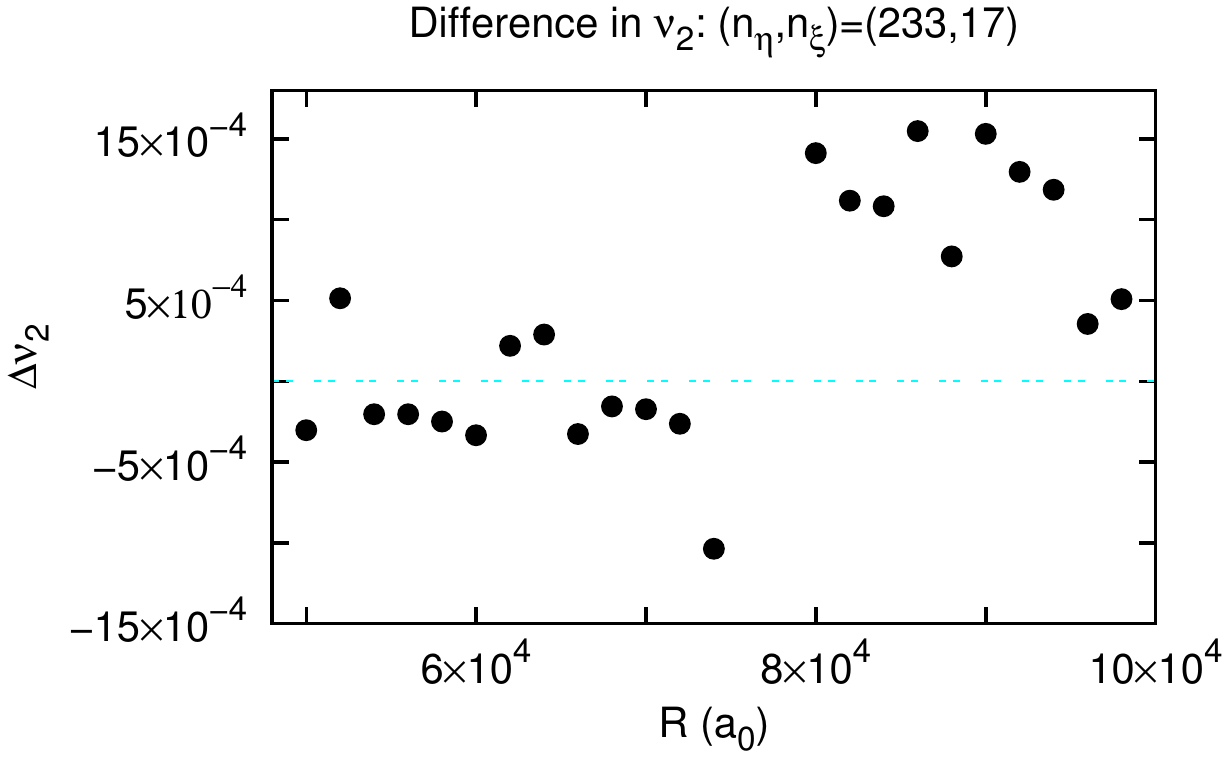}}

\subfigure[]{\includegraphics[width=0.95\columnwidth]{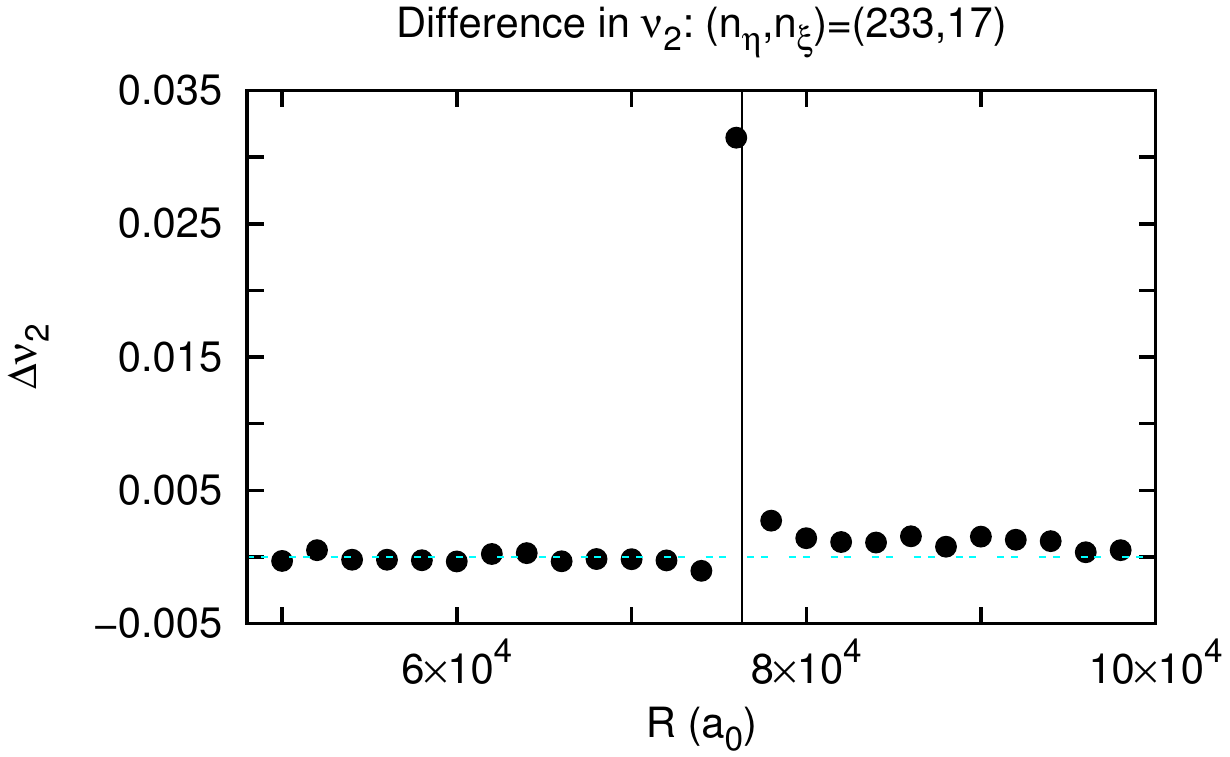}}
\caption{Comparison between quantum (solid lines) and semiclassical (dashed lines) energies for selected high-lying states of H$_2^+$. Panel~(a) shows that the semiclassical and quantum PECs are very similar. Panel~(b) presents the error in $\nu_2$ for the state $(n_{\eta},n_{\xi})=(233,17)$ away from the barrier top, while Panel~(c) shows the error increase near the barrier top (vertical line) and quickly drop back down.} 
\label{fig:comparepec2}
\end{figure}}
{
\begin{figure}[!ht]
\subfigure[]{\includegraphics[width=5.5in]{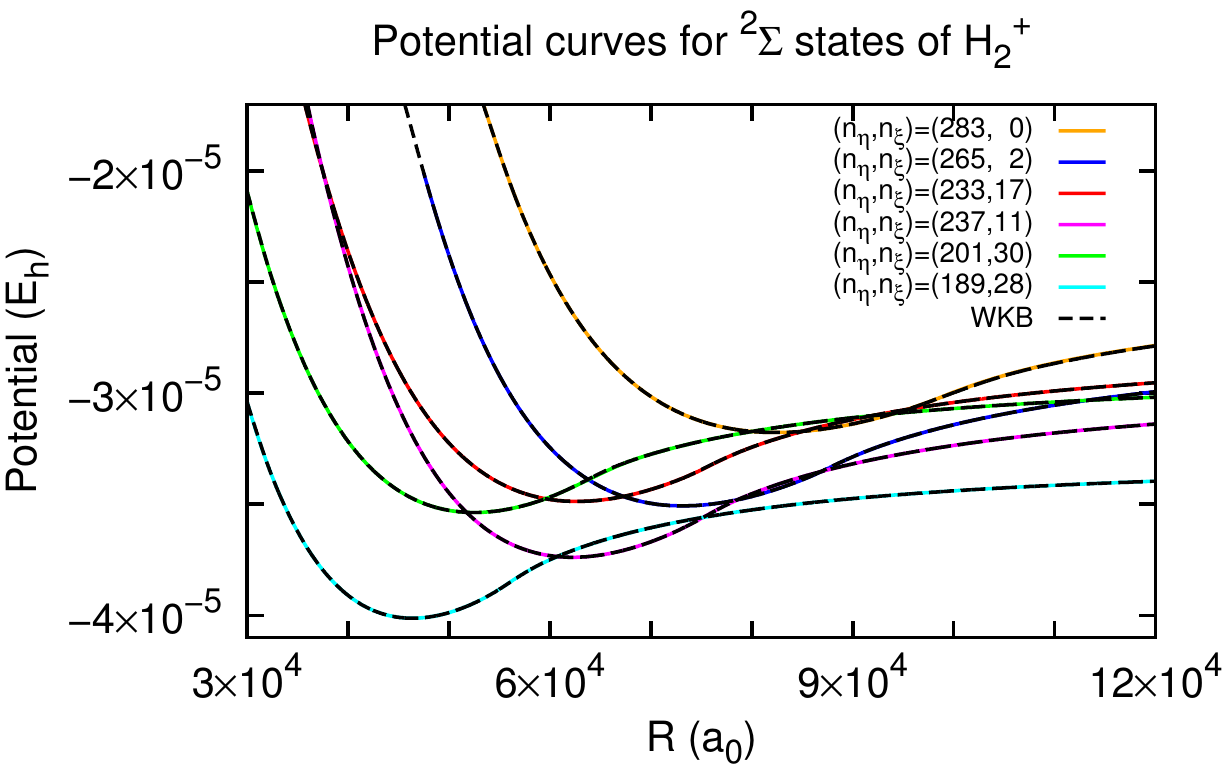}}

\subfigure[]{\includegraphics[width=3.1in]{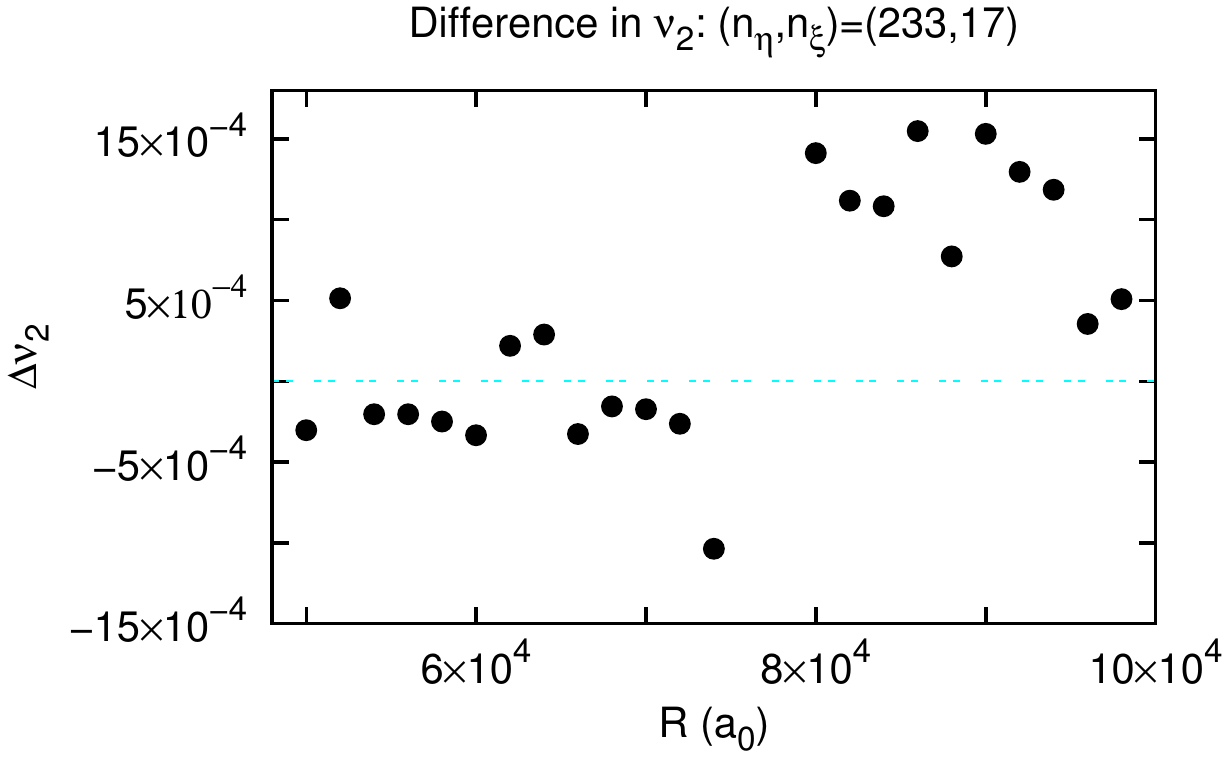}}
\subfigure[]{\includegraphics[width=3.1in]{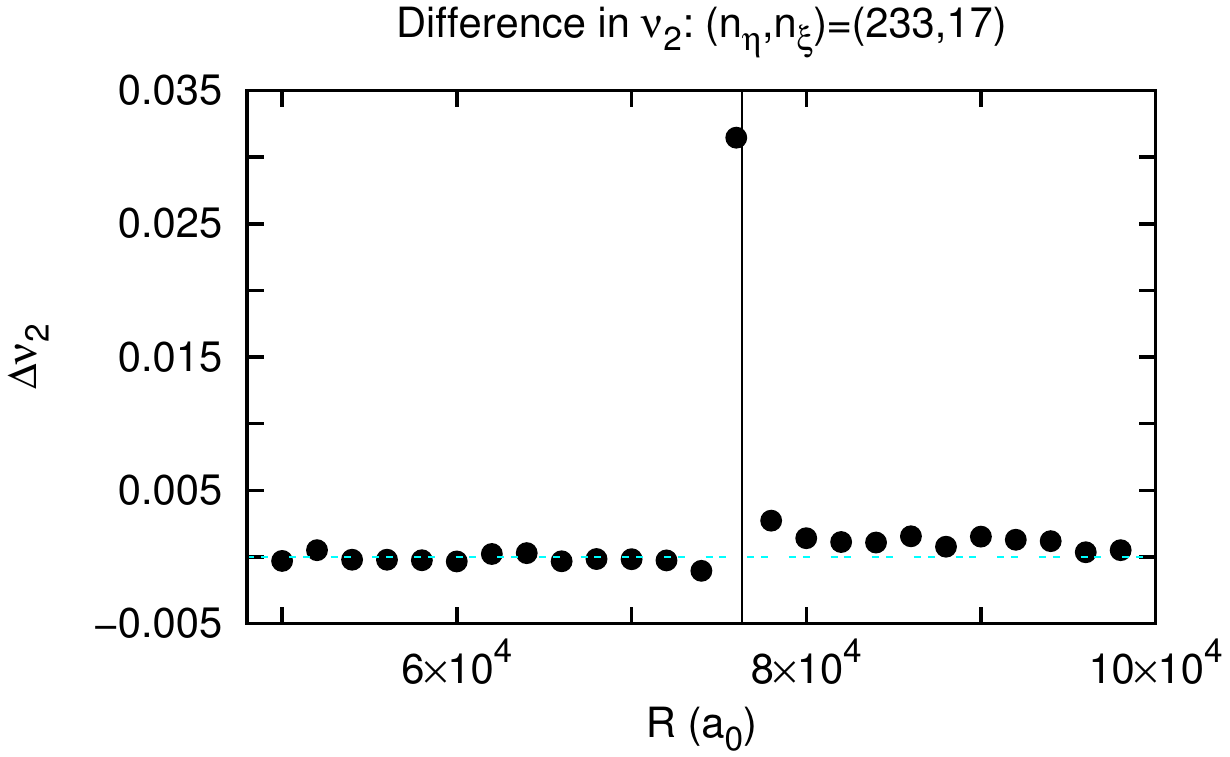}}
\caption{Comparison between quantum (solid lines) and semiclassical (dashed lines) energies for selected high-lying states of H$_2^+$. Panel~(a) shows that the semiclassical and quantum PECs are very similar. Panel~(b) presents the error in $\nu_2$ for the state $(n_{\eta},n_{\xi})=(233,17)$ away from the barrier top, while Panel~(c) shows the error increase near the barrier top (vertical line) and quickly drop back down.} 
\label{fig:comparepec2}
\end{figure}
}
\ifthenelse{\boolean{twocolumn}}{
\begin{table}[htb]
\caption{Exact and Semiclassical Spectroscopic Constants for the States Presented in Fig.~\ref{fig:comparepec2}.}
\label{tab:spectconst}
\centering
\resizebox{0.9\columnwidth}{!}{%
\begin{tabular}{cc|ccc} \hline
 \multicolumn{2}{c|} {state $(n_{\eta},n_{\xi})$}  & R$_{\mathrm{e}}$ $(a_0)$ & $\omega_{\mathrm{e}}$\,{(MHz)} & $D_{\mathrm{e}}$\,(GHz)\\ \hline
\multirow{2}{*}  {(189,28)}  & {QM}  & 46698   & 23.16 &  181.367  \\
&{WKB} & 46698  & 23.16 &  181.368  \\ \hline
\multirow{2}{*}  {(201,30)}  &{QM} & 52646 & 18.50 & 157.150   \\
&{WKB} & 52646 & 18.50 & 157.151   \\ \hline
\multirow{2}{*}  {(237,11)} &{QM} & 62245 & 18.14 & 154.426   \\
&{WKB} & 62246 & 18.13 & 154.427   \\ \hline
\multirow{2}{*}  {(233,17)} &{QM} & 62709 & 16.82 & 147.919   \\
&{WKB} & 62708 & 16.82 & 147.920   \\ \hline
\multirow{2}{*}  {(265,2)} &{QM} & 73152 & 15.75 & 137.397   \\
&{WKB} & 73154 & 15.70 & 137.398   \\ \hline
\multirow{2}{*}  {(283,0)} &{QM} & 82444 & 13.52 & 123.805   \\
&{WKB} & 82443 & 13.51 & 123.805   \\ \hline
\end{tabular}}
\end{table}}   
{
\begin{table}[htb]
\caption{Exact and Semiclassical Spectroscopic Constants for the States Presented in Fig.~\ref{fig:comparepec2}.}
\label{tab:spectconst}
\centering
\begin{tabular}{cc|ccc} \hline
 \multicolumn{2}{c|} {state $(n_{\eta},n_{\xi})$}  & R$_{\mathrm{e}}$ $(a_0)$ & $\omega_{\mathrm{e}}$\,{(MHz)} & $D_{\mathrm{e}}$\,(GHz)\\ \hline
\multirow{2}{*}  {(189,28)}  & {QM}  & 46698   & 23.16 &  181.367  \\
&{WKB} & 46698  & 23.16 &  181.368  \\ \hline
\multirow{2}{*}  {(201,30)}  &{QM} & 52646 & 18.50 & 157.150   \\
&{WKB} & 52646 & 18.50 & 157.151   \\ \hline
\multirow{2}{*}  {(237,11)} &{QM} & 62245 & 18.14 & 154.426   \\
&{WKB} & 62246 & 18.13 & 154.427   \\ \hline
\multirow{2}{*}  {(233,17)} &{QM} & 62709 & 16.82 & 147.919   \\
&{WKB} & 62708 & 16.82 & 147.920   \\ \hline
\multirow{2}{*}  {(265,2)} &{QM} & 73152 & 15.75 & 137.397   \\
&{WKB} & 73154 & 15.70 & 137.398   \\ \hline
\multirow{2}{*}  {(283,0)} &{QM} & 82444 & 13.52 & 123.805   \\
&{WKB} & 82443 & 13.51 & 123.805   \\ \hline
\end{tabular}
\end{table}} 

\subsection{Wave functions}
Figure~\ref{fig:comparewf} illustrates the agreement of the semiclassical and quantum mechanical $\eta$ wave functions for large $R$ near 4 $\mu${m}; the PECs for these states are shown in Fig.~\ref{fig:comparepec2}. 
\ifthenelse{\boolean{twocolumn}}{
\begin{figure}[!ht]
\subfigure[]{\includegraphics[width=0.95\columnwidth]{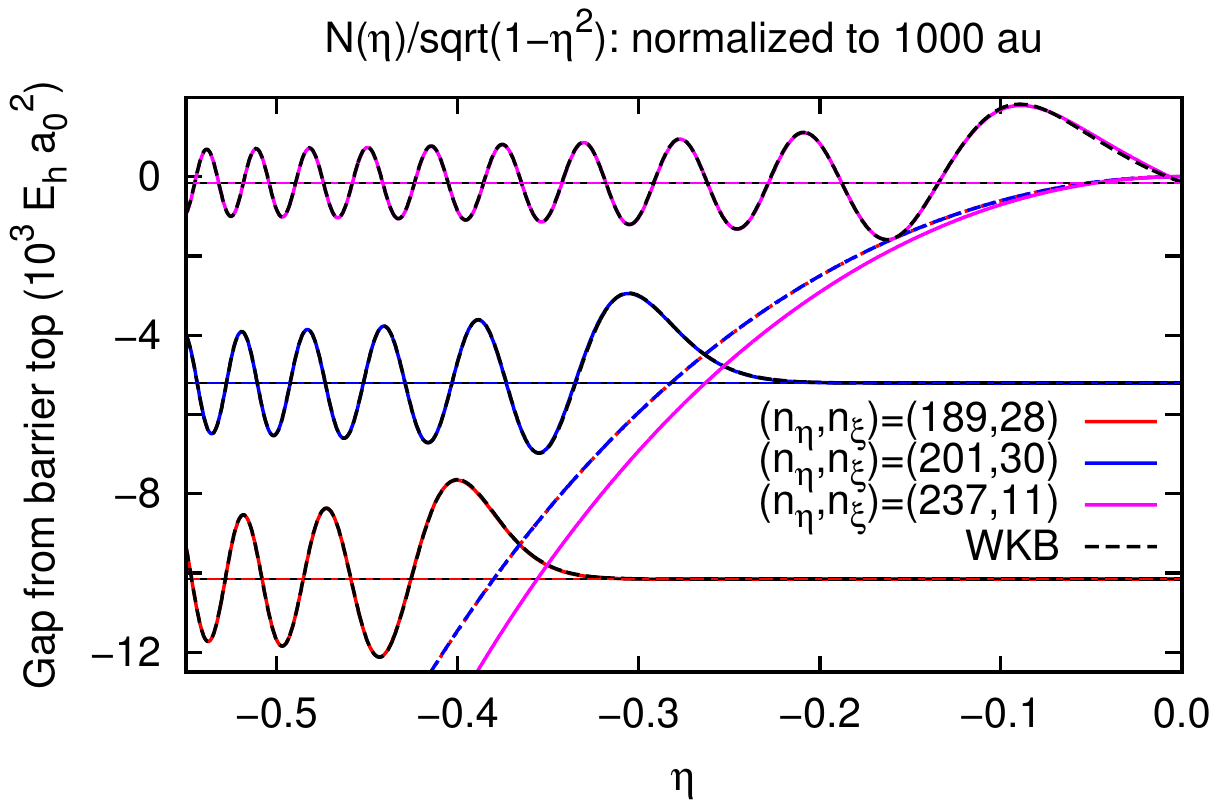}}

\subfigure[]{\includegraphics[width=0.95\columnwidth]{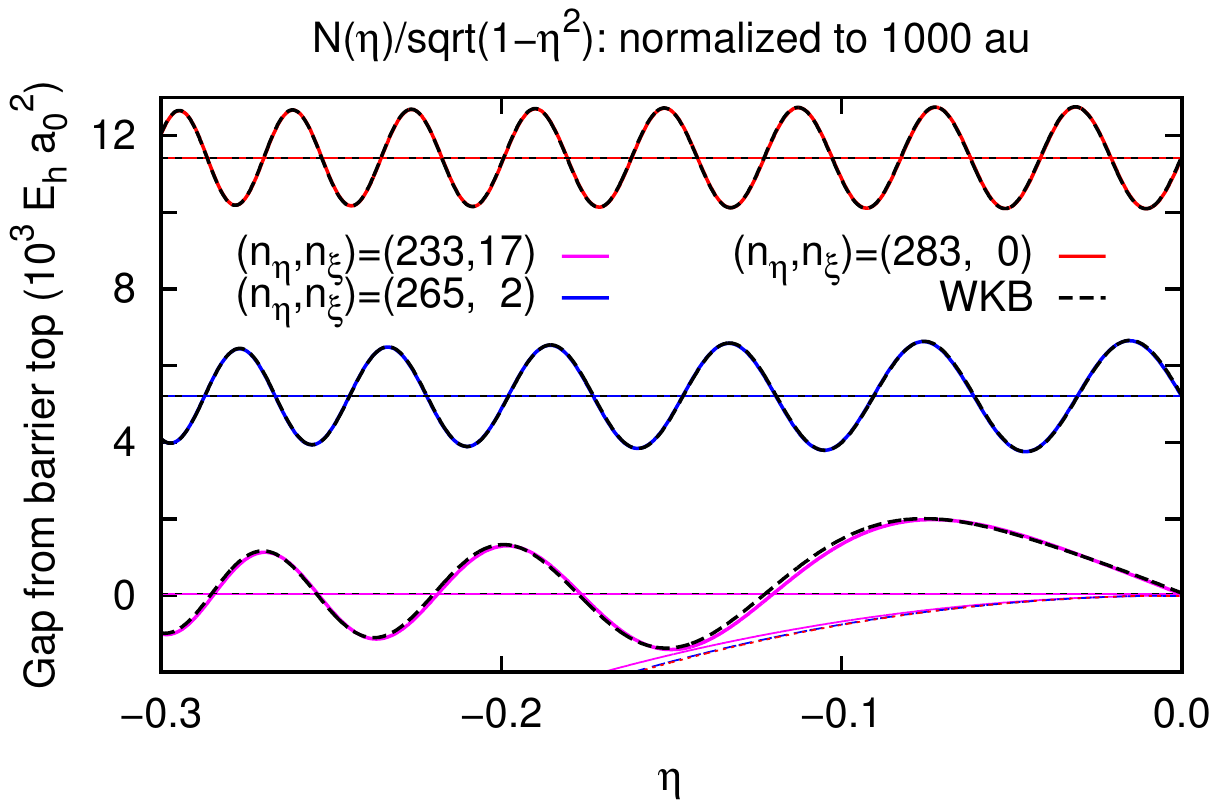}}
\caption{Comparison between quantum (solid lines) and semiclassical (dashed lines) wave functions $N(\eta)/\sqrt{1-\eta^2}$ at $R=76000$~$a_0$ below (Panel~(a)) and above the barrier (Panel~(b)). The PECs for these states were shown in Fig.~\ref{fig:comparepec2}. The vertical axis gives the gap from the top of the barrier to the ``energy'' $(\gamma^2/2)$ of the state, and the wave functions are normalized to 1000 atomic units for visibility. For the states below the barrier, we have put an Airy patch on the wave function near the classical turning point. The agreement improves for states farther from the barrier top. } 
\label{fig:comparewf}
\end{figure}}
{\begin{figure}[!ht]
\subfigure[]{\includegraphics[width=4.5in]{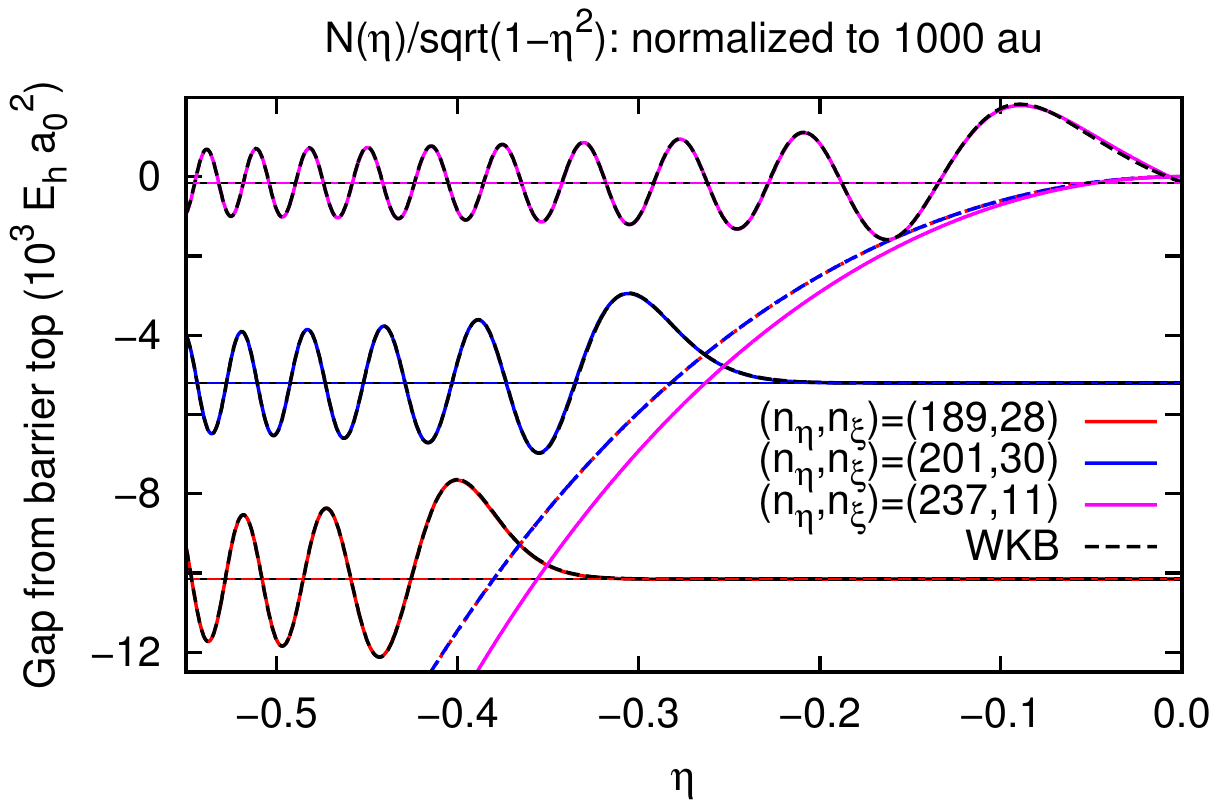}}

\subfigure[]{\includegraphics[width=4.5in]{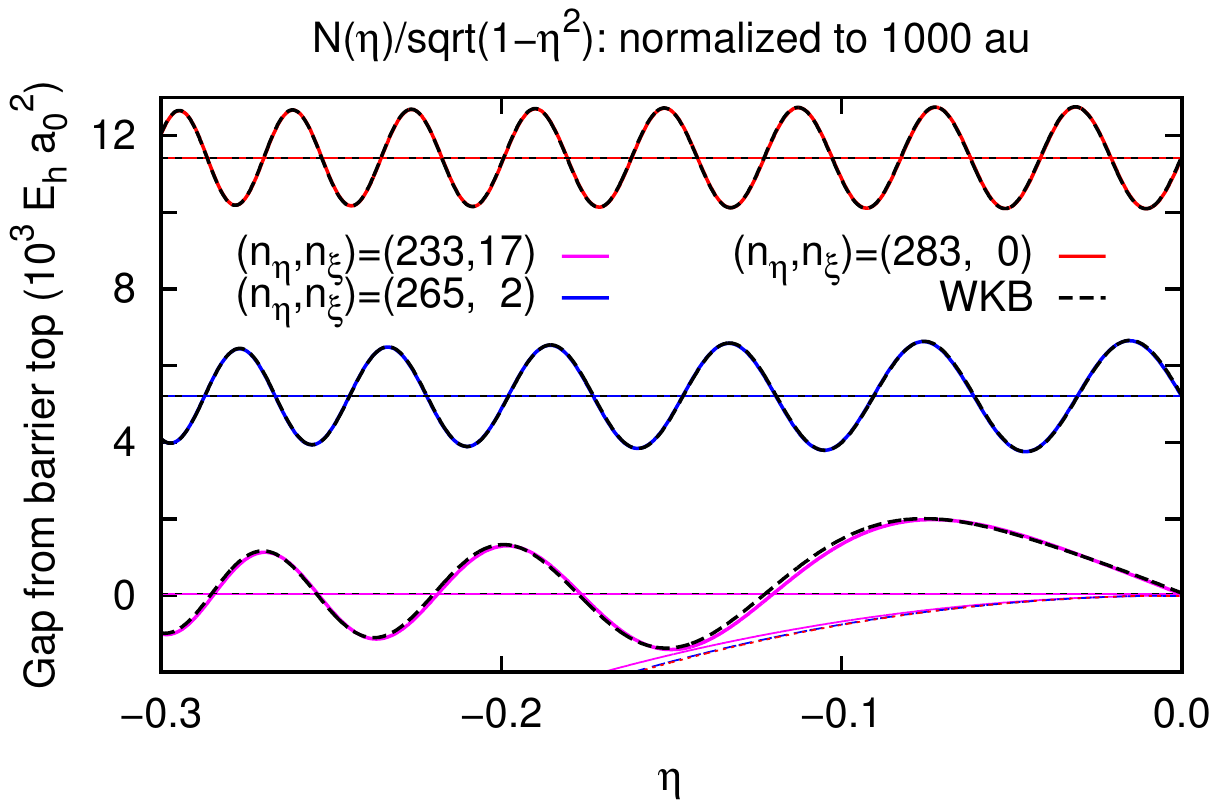}}
\caption{Comparison between quantum (solid lines) and semiclassical (dashed lines) wave functions $N(\eta)/\sqrt{1-\eta^2}$ at $R=76000$~$a_0$ below (Panel~(a)) and above the barrier (Panel~(b)). The PECs for these states were shown in Fig.~\ref{fig:comparepec2}. The vertical axis gives the gap from the top of the barrier to the ``energy'' $(\gamma^2/2)$ of the state, and the wave functions are normalized to 1000 atomic units for visibility. For the states below the barrier, we have put an Airy patch on the wave function near the classical turning point. The agreement improves for states farther from the barrier top. } 
\label{fig:comparewf}
\end{figure}}
For the states below the barrier shown in Panel~(a) of Fig.~\ref{fig:comparewf}, we used an Airy patch (Eq.~\ref{eq:AiryEta}) so that the wave function in $\eta$ would match the exact wave function better near the classical turning point. We connected the WKB wavefunction in the classical region to the Airy patch using a taut spline routine\cite{Spline}; this ensured that both the derivative and value of the wave function was continuous and that we did not introduce extra oscillations. The Airy patch is inadequate near the top of the barrier, where the potential is better approximated by a parabola, which is the case for the state $(237,11)$ in Panel~(a) of Fig.~\ref{fig:comparewf}. This is where the uniform approach utilized by Strand and Reinhardt \cite{Reinhardt} is most appropriate. 

The semiclassical excited state wave functions are used to compute dipole matrix elements with the ground state. The calculations use the quantum mechanical ground state wavefunction for the dipole matrix elements, so that any differences from their exact values reflect only errors in the upper state ungerade wave functions. Some results are given in Table~\ref{tab:matels}. Below the barrier, the dipole matrix elements in Table~\ref{tab:matels} differ by less than 5\%. Above the barrier, they differ by less than 30\%. Although the increased discrepancy above the barrier is likely due to the small values $n_{\xi}=0$ and 2, it could also be the fact that below the barrier the $\eta$ wave functions were splined to the more exact Airy functions, leading to a more accurate normalization constant. Even for $n_{\xi}=0$, the agreement between the quantum and semiclassical matrix elements is very good. 
\ifthenelse{\boolean{twocolumn}}{
\begin{table}[htb]
\caption{Comparison between Dipole Matrix Elements ``DMEs'' Calculated by Using Exact and Semiclassical Wave Functions for the States of H$^+_2$ {Shown in Fig.~\ref{fig:comparewf}} and Taken with Respect to the Quantum Gerade Ground State.}
\label{tab:matels}
\centering
\resizebox{0.7\columnwidth}{!}{%
\begin{tabular}{cc|c} \hline
\multicolumn{2}{c|} {state $(n_{\eta},n_{\xi})$}  & {DME} ($10^{-5}$ a.u.)   \\ \hline 
\multirow{2}{*} {(189,28)} & {QM} & $7.685$   \\
& {WKB} & $7.804$   \\ \hline 
\multirow{2}{*} {(201,30)} &{QM} & $6.459$ \\
&{WKB}& $6.620$ \\ \hline
\multirow{2}{*} {(237,11)} &{QM}& $7.441$  \\
&{WKB}& $7.692$    \\ \hline
\multirow{2}{*} {(233,17)} &{QM}& $6.431$  \\
&{WKB}& $6.794$ \\ \hline
\multirow{2}{*} {(265,2)} &{QM}& $11.58$   \\
&{WKB}& $14.07$  \\ \hline
\multirow{2}{*} {(283,0)} &{QM}& $12.71$  \\
&{WKB}& $18.53$  \\ \hline
\end{tabular}}
\end{table}}   
{\begin{table}[htb]
\caption{Comparison between dipole matrix elements ``DMEs'' calculated by using exact and semiclassical wave functions for the states of H$^+_2$ {shown in Fig.~\ref{fig:comparewf}}. The DMEs were taken with respect to the quantum gerade ground state.}
\label{tab:matels}
\centering
\begin{tabular}{cc|c} \hline
\multicolumn{2}{c|} {state $(n_{\eta},n_{\xi})$}  & {DME} ($10^{-5}$ a.u.)  \\ \hline 
\multirow{2}{*} {(189,28)} & {QM} & $7.685$   \\
& {WKB} & $7.804$   \\ \hline 
\multirow{2}{*} {(201,30)} &{QM} & $6.459$ \\
&{WKB}& $6.620$ \\ \hline
\multirow{2}{*} {(237,11)} &{QM}& $7.441$  \\
&{WKB}& $7.692$    \\ \hline
\multirow{2}{*} {(233,17)} &{QM}& $6.431$  \\
&{WKB}& $6.794$ \\ \hline
\multirow{2}{*} {(265,2)} &{QM}& $11.58$   \\
&{WKB}& $14.07$  \\ \hline
\multirow{2}{*} {(283,0)} &{QM}& $12.71$  \\
&{WKB}& $18.53$  \\ \hline
\end{tabular}
\end{table}}

Figure~\ref{fig:comparephaseeta} compares the phases of the wave functions $N(\eta)$ for the states presented in Figs.~\ref{fig:comparepec2} and \ref{fig:comparewf}. The quantum phases (solid lines) are determined through Milne's phase-amplitude procedure, with the phase defined as:
\begin{equation}
\phi_{\mathrm{Milne}}(\eta)=\int_{-1}^{\eta} \alpha^{-2}(\eta')\,d\eta'. \label{eq:Milneph}
\end{equation}
The dashed lines show the semiclassical phases in the classical region, 
\begin{equation}
\phi_{\mathrm{WKB}}(\eta)=\int_{-1}^{\eta} k_{\mathrm{L}}(\eta')\,d\eta' + \pi/4,\hspace{10mm} \eta<-\eta_{\mathrm{c}} \label{eq:WKBph}
\end{equation}
where $\eta_{\mathrm{c}}=0$ for states above the barrier. The quantum (Eq.~\ref{eq:Milneph}) and semiclassical (Eq.~\ref{eq:WKBph}) phases are most different for low values of $\eta$ where the quasimomentum varies rapidly and the WKB approximation is less accurate. This small $\eta$ region is highlighted in Panel~(b). The discrepancies are still small, and over the whole range of $\eta$ the quantum and semiclassical results differ by less than 0.1\%. 
\ifthenelse{\boolean{twocolumn}}{
\begin{figure}[!ht]
\subfigure[]{\includegraphics[width=0.95\columnwidth]{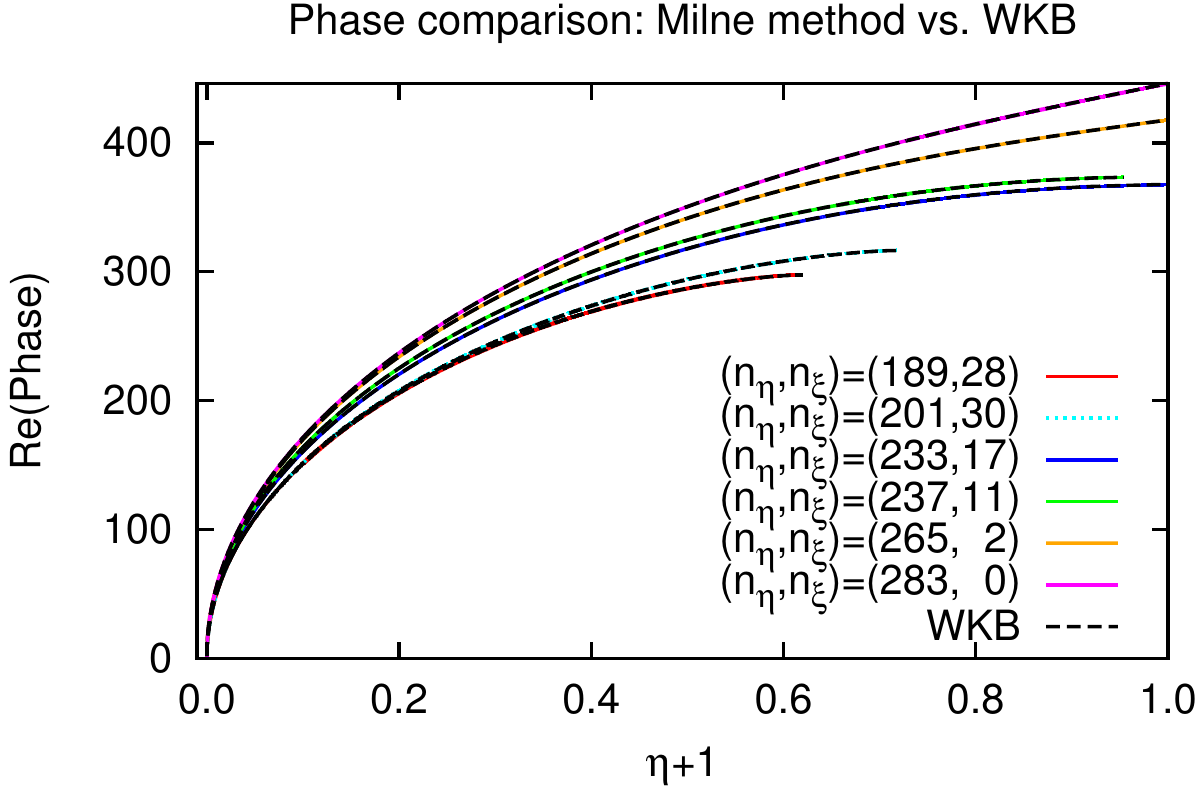}}
\subfigure[]{\includegraphics[width=0.95\columnwidth]{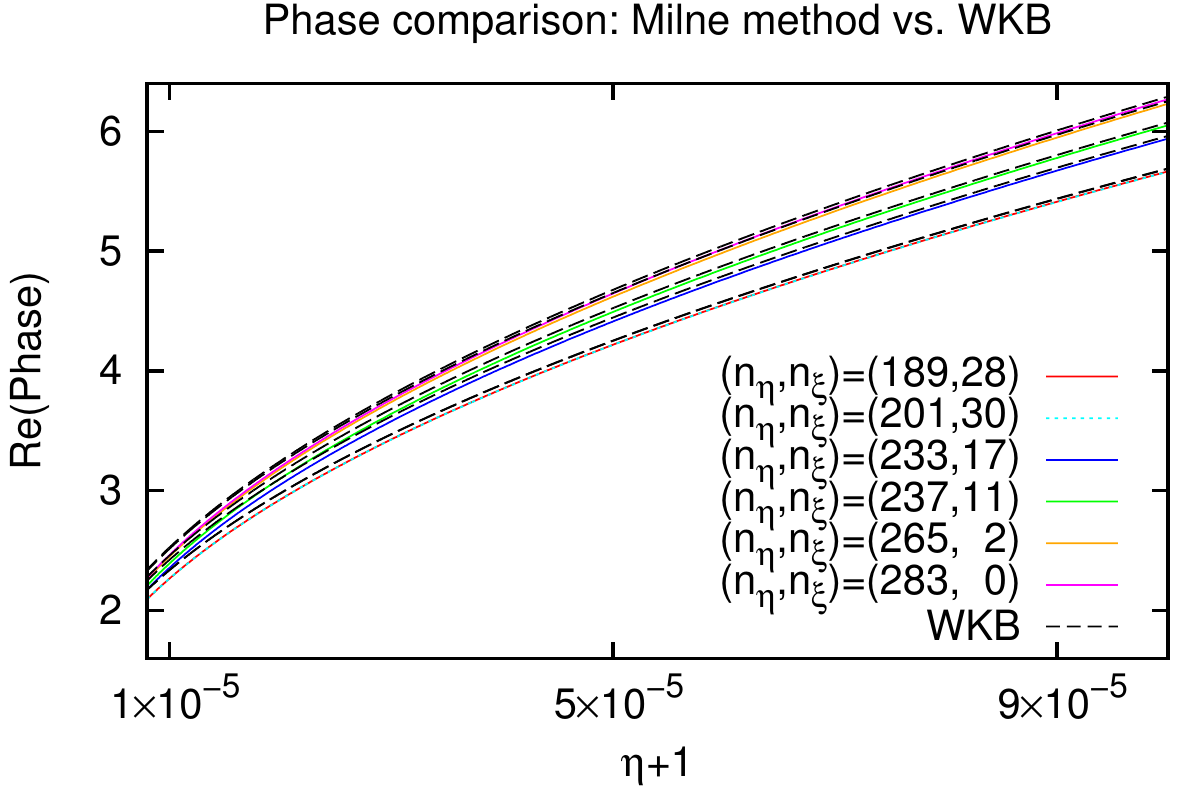}}
\caption{Comparison between Milne (solid lines) and semiclassical (dashed lines) phases of the wave functions $N(\eta)$ for the states presented in Figs.~\ref{fig:comparepec2} and \ref{fig:comparewf} at $R=76000$~$a_0$. The phases are defined by Eqs.~\ref{eq:Milneph} and \ref{eq:WKBph}. Panel~(a) shows the comparison over the range $-1\leq\eta\leq -\eta_{\mathrm{c}}$, while Panel~(b) highlights the small $\eta$ portion.} 
\label{fig:comparephaseeta}
\end{figure}}
{\begin{figure}[!ht]
\subfigure[]{\includegraphics[width=5.0in]{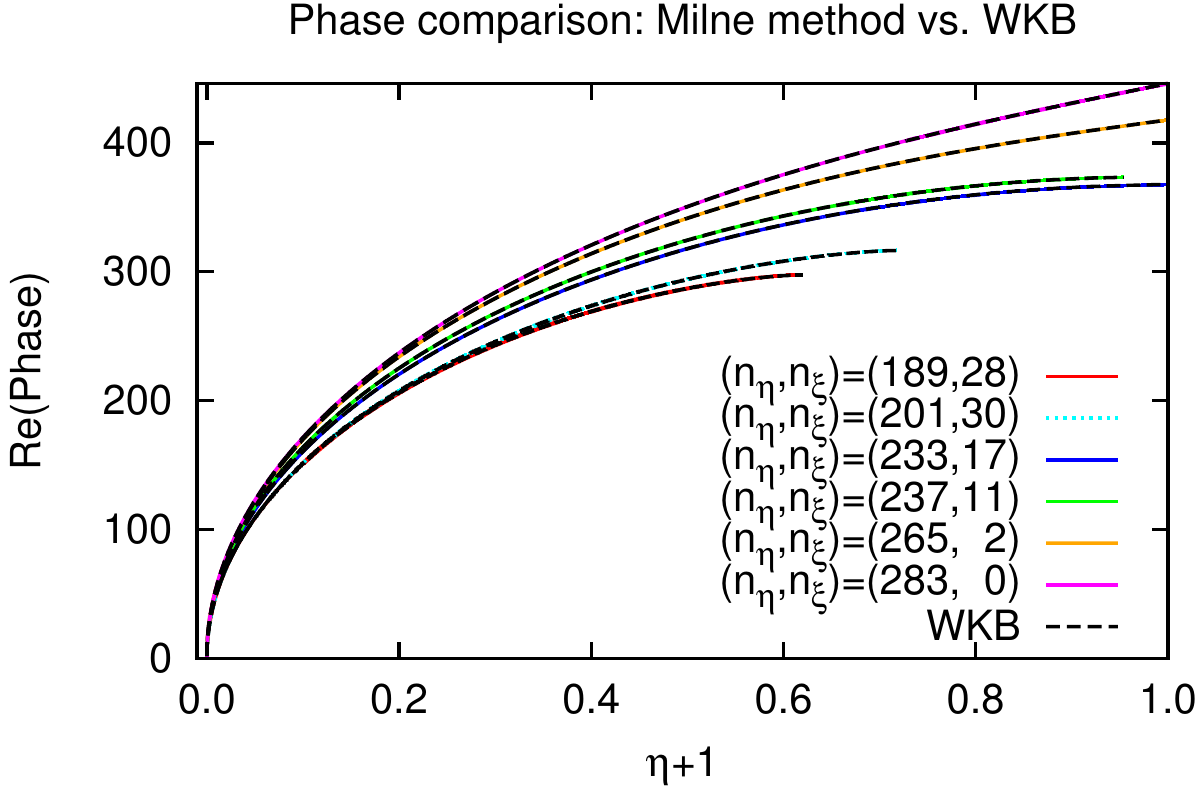}}
\subfigure[]{\includegraphics[width=5.0in]{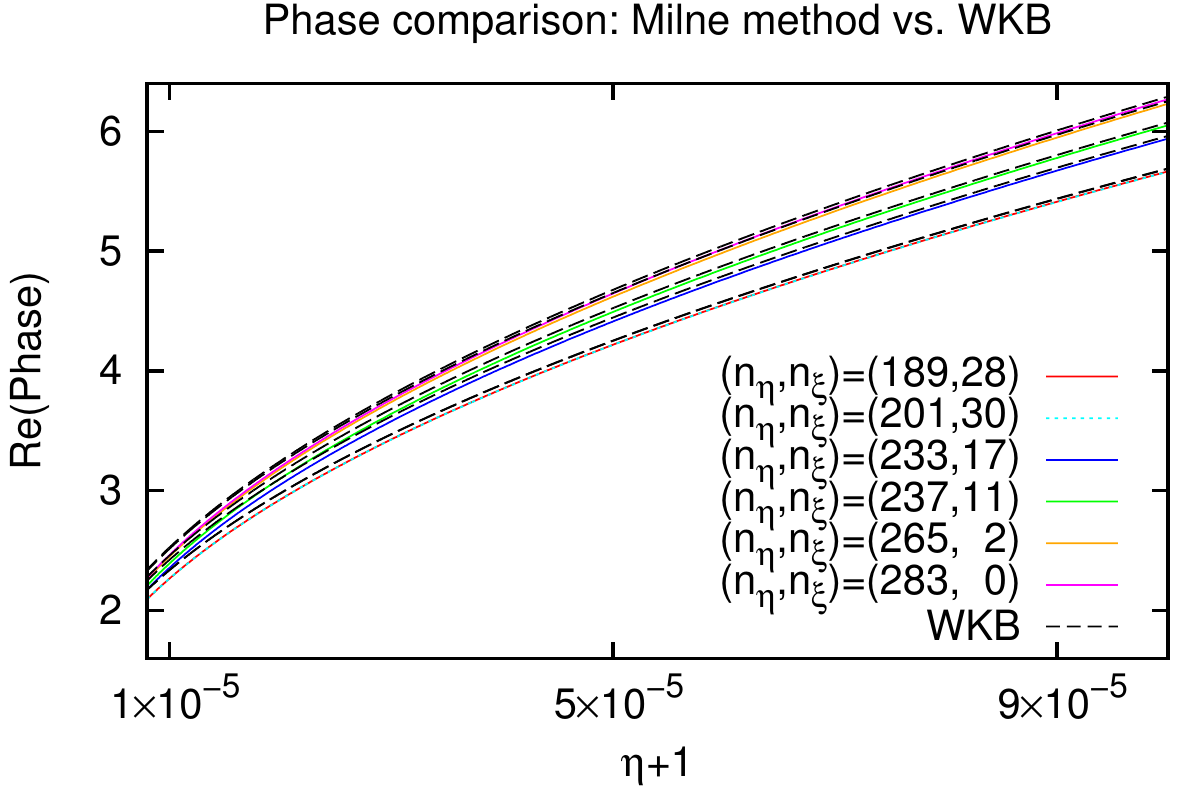}}
\caption{Comparison between Milne (solid lines) and semiclassical (dashed lines) phases of the wave functions $N(\eta)$ for the states presented in Figs.~\ref{fig:comparepec2} and \ref{fig:comparewf} at $R=76000$~$a_0$. The phases are defined by Eqs.~\ref{eq:Milneph} and \ref{eq:WKBph}. Panel~(a) shows the comparison over the range $-1\leq\eta\leq -\eta_{\mathrm{c}}$, while Panel~(b) highlights the small $\eta$ portion.} 
\label{fig:comparephaseeta}
\end{figure}}

Figure~\ref{fig:comparephasexi} compares the phases of the wave functions $X(\xi)$, defined as
\begin{eqnarray}
\phi_{\mathrm{Milne}}(\xi)&=&\int_1^{\xi} \alpha^{-2}(\xi')\,d\xi'. \label{eq:Milnephxi} \\
\phi_{\mathrm{WKB}}(\xi)&=&\int_{\xi_{\mathrm{c}1}}^{\xi} k_{\mathrm{L}}(\xi')\,d\xi' + \pi/4\hspace{10mm} \xi<\xi_{\mathrm{c}2}. \label{eq:WKBphxi}
\end{eqnarray}
Panel~(b) shows that, as with the phases of $N(\eta)$, the quantum (solid lines) and WKB (dashed lines) values differ most when $\xi$ is small. The deviation is larger for wave functions with a smaller node number.
\ifthenelse{\boolean{twocolumn}}{
\begin{figure}[!ht]
\subfigure[]{\includegraphics[width=0.95\columnwidth]{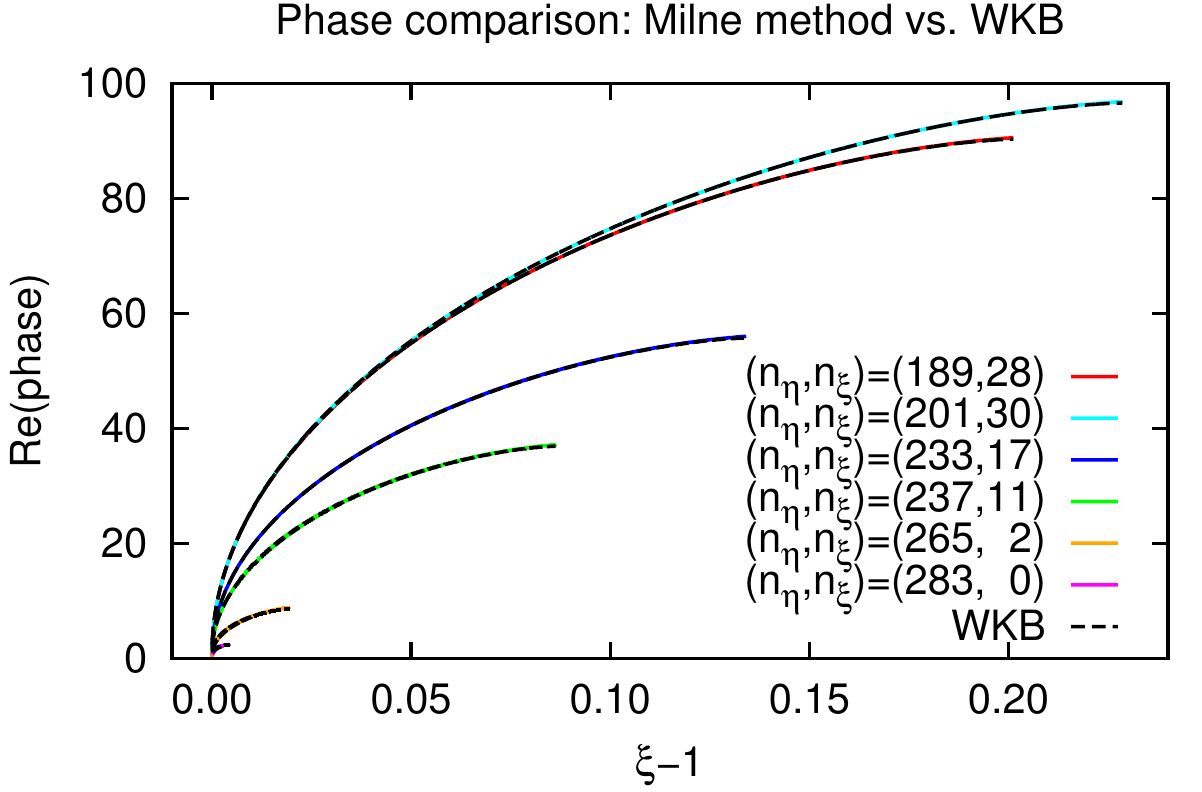}}
\subfigure[]{\includegraphics[width=0.95\columnwidth]{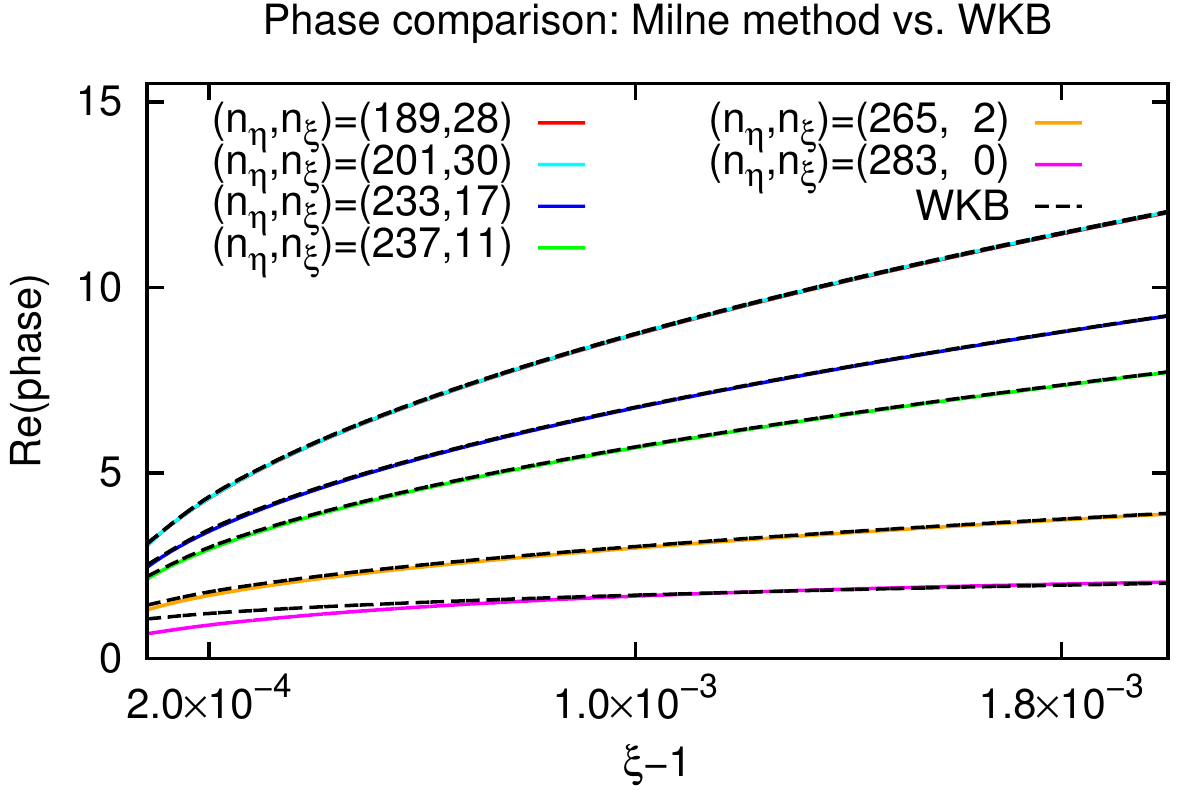}}
\caption{Comparison between Milne (solid lines) and semiclassical (dashed lines) phases of the wave functions $X(\xi)$ for the states presented in Figs.~\ref{fig:comparepec2} and \ref{fig:comparewf} at $R=76000$~$a_0$. The phases are defined by Eqs.~\ref{eq:Milnephxi} and \ref{eq:WKBphxi}. Panel~(a) shows the comparison for a range of $\xi$ extending to the classical turning points of the potentials, while Panel~(b) shows the comparison for small $\xi$.} 
\label{fig:comparephasexi}
\end{figure}}
{\begin{figure}[!ht]
\subfigure[]{\includegraphics[width=5.0in]{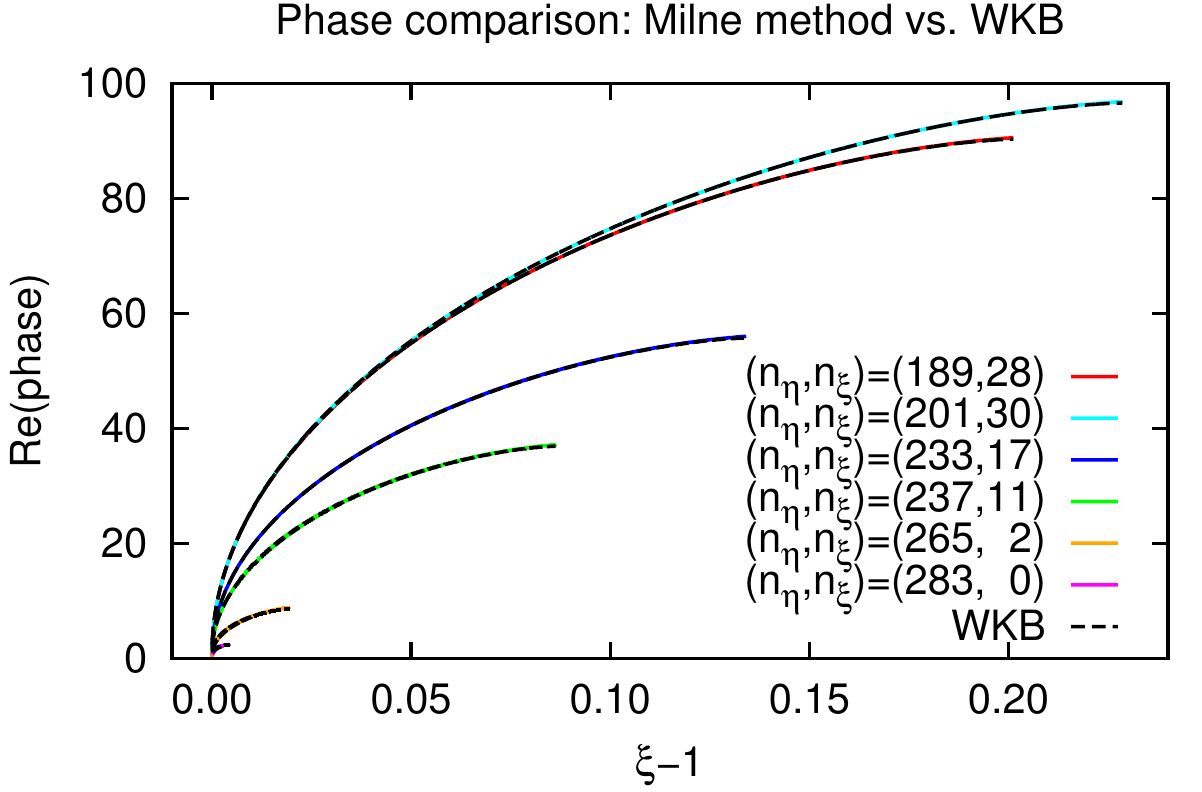}}
\subfigure[]{\includegraphics[width=5.0in]{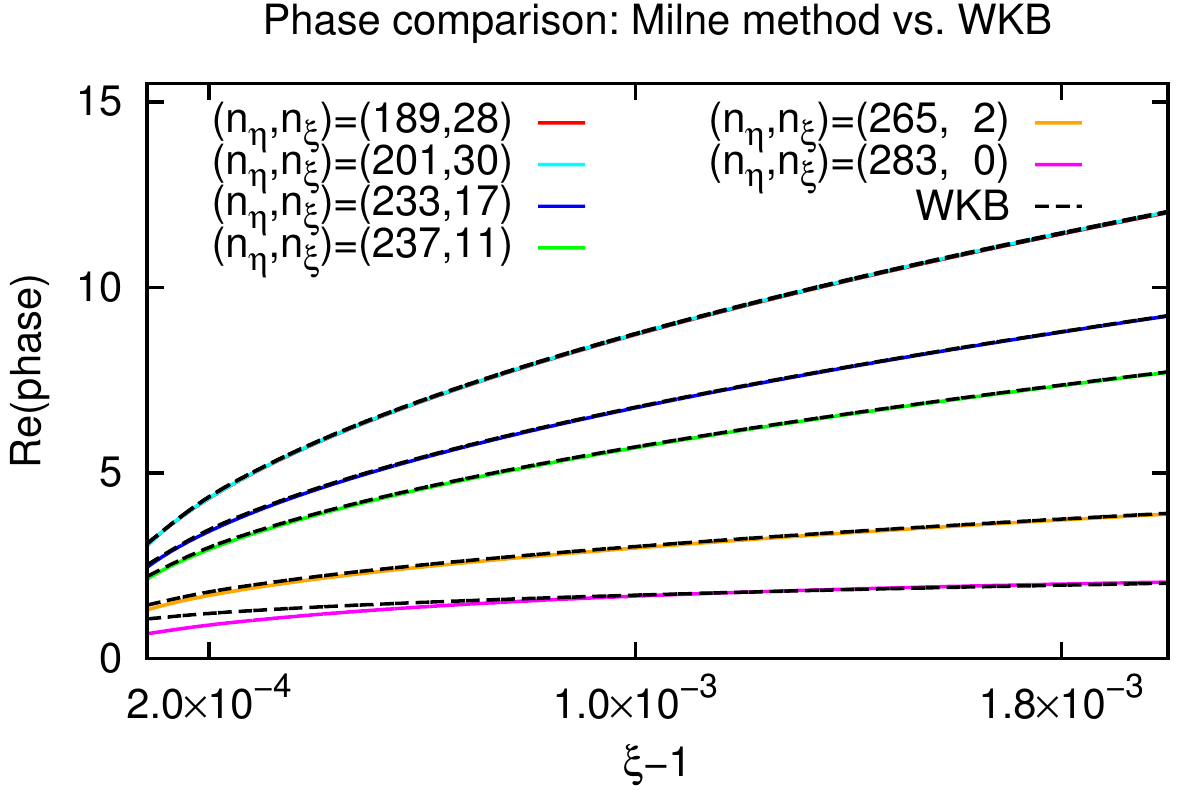}}
\caption{Comparison between Milne (solid lines) and semiclassical (dashed lines) phases of the wave functions $X(\xi)$ for the states presented in Figs.~\ref{fig:comparepec2} and \ref{fig:comparewf} at $R=76000$~$a_0$. The phases are defined by Eqs.~\ref{eq:Milnephxi} and \ref{eq:WKBphxi}. Panel~(a) shows the comparison for a range of $\xi$ extending to the classical turning points of the potentials, while Panel~(b) shows the comparison for small $\xi$.} 
\label{fig:comparephasexi}
\end{figure}}

Figure~\ref{fig:comparekxi} presents a typical comparison between the WKB wave number $k_{\mathrm{L}}(\xi)$ and the Milne $\alpha^{-2}(\xi)$. The dotted line shows the percent difference between the two, which increases as the WKB approximation breaks down. Approximately, the WKB assumptions are satisfied where 
\begin{equation}
\left|\frac{d}{d\xi}\frac{1}{k_{\mathrm{L}}(\xi)}\right|<<1 \label{eq:WKBcond}
\end{equation}
The region for which the left hand side of Eq.~\ref{eq:WKBcond} is less than 0.1 is bracketed by the two vertical lines in Fig.~\ref{fig:comparekxi}, with the vertical line on the left at approximately $\xi=1.003$ and the vertical line on the right at approximately $\xi=1.094$. The percent difference between $k_{\mathrm{L}}(\xi)$ and $\alpha^{-2}(\xi)$ is less than 0.5 within this region. For small $\xi<1.0030$, $k_{\mathrm{L}}(\xi)$ varies more and more rapidly before becoming singular at $\xi=1$, so that the left-hand side of Eq.~\ref{eq:WKBcond} increases as $\xi$ decreases, and the percent difference between $k_{\mathrm{L}}(\xi)$ and $\alpha^{-2}(\xi)$ grows to be on the order of 100. For $\xi>1.094$, near the classical turning point of the potential, the percent difference again grows to be quite large.
\ifthenelse{\boolean{twocolumn}}{
\begin{figure}[!ht]
{\includegraphics[width=0.95\columnwidth]{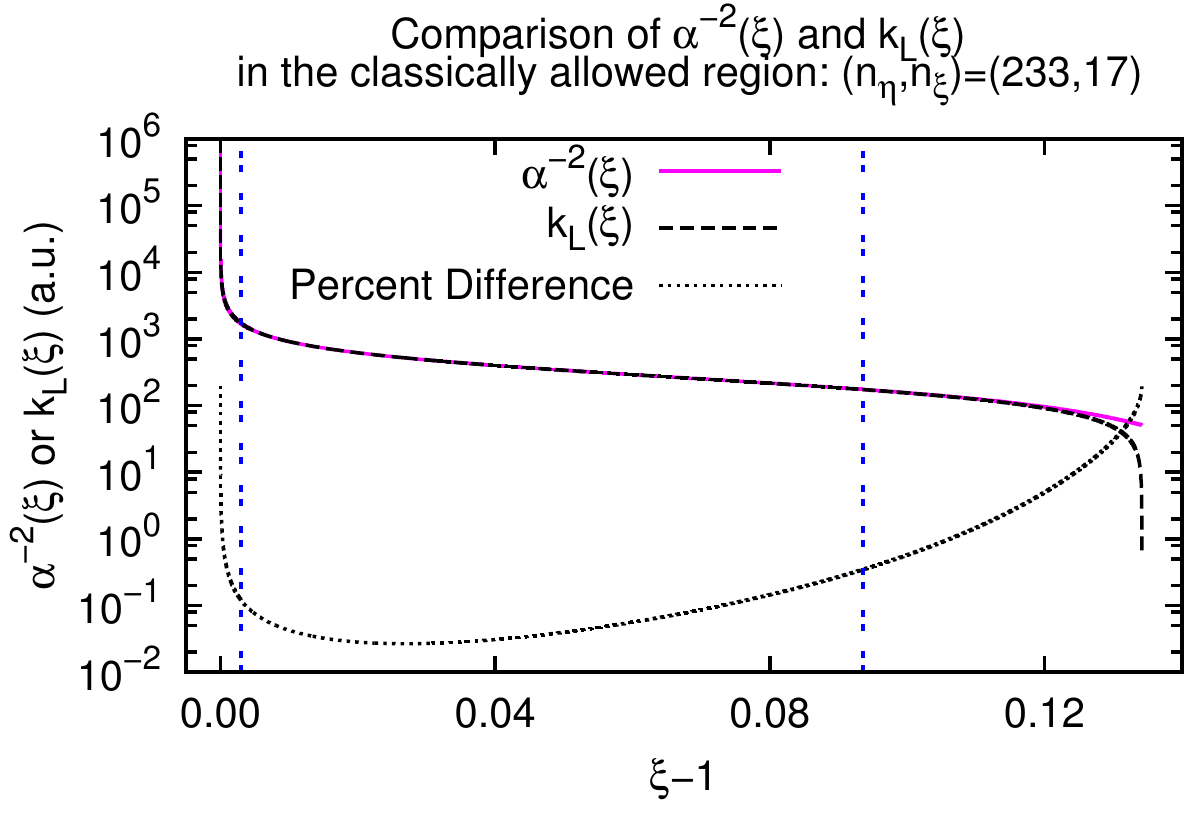}}
\caption{Comparison between the Milne $\alpha^{-2}(\xi)$ (solid line) and the WKB quasimomentum $k_{\mathrm{L}}(\xi)$ (dashed line) for the state $(n_{\eta},n_{\xi})=(233,17)$ at $R=76000~a_0$. The percent difference between the two quantities, defined as $200\times(\alpha^{-2}(\xi)-k_{\mathrm{L}}(\xi))/(\alpha^{-2}(\xi)+k_{\mathrm{L}}(\xi))$, is shown as a dotted line. The two vertical lines bound the region where the left hand side of Eq.~\ref{eq:WKBcond} is less than or equal to 0.1.} 
\label{fig:comparekxi}
\end{figure}}
{\begin{figure}[!ht]
{\includegraphics[width=5.0in]{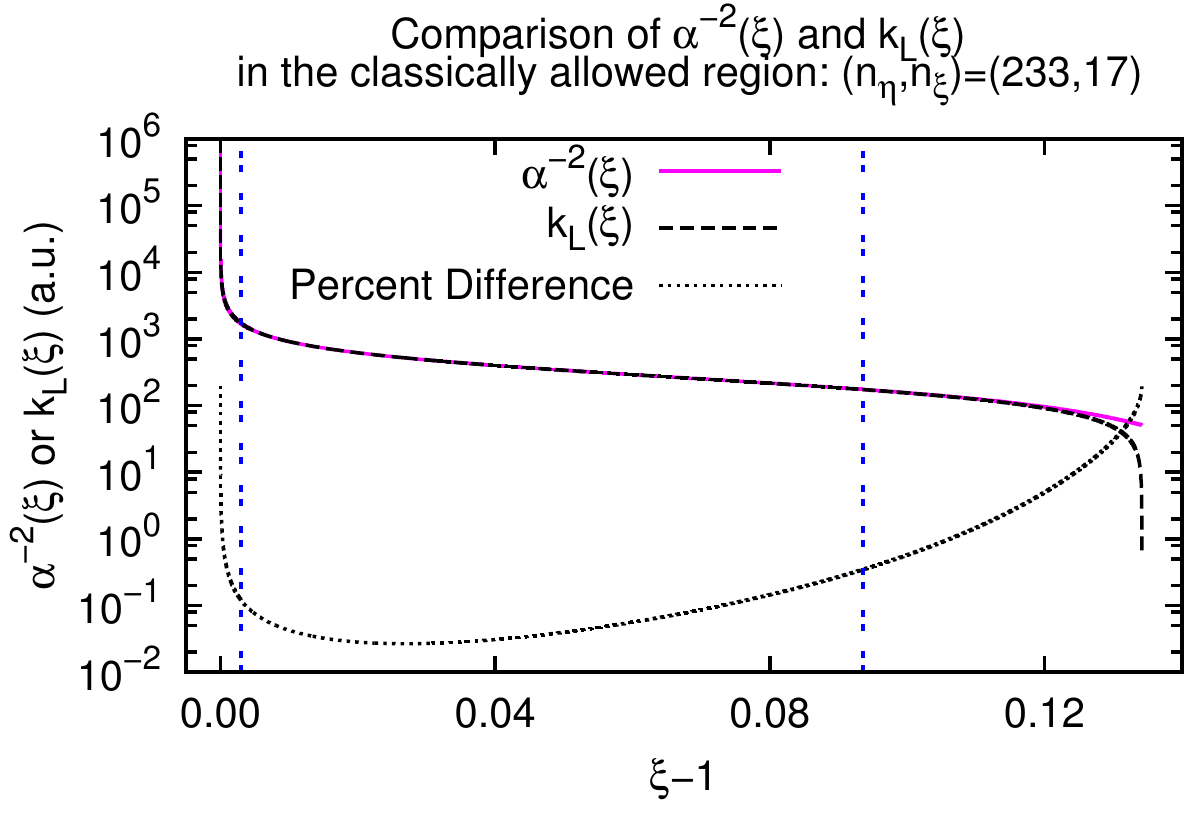}}
\caption{Comparison between the Milne $\alpha^{-2}(\xi)$ (solid line) and the WKB quasimomentum $k_{\mathrm{L}}(\xi)$ (dashed line) for the state $(n_{\eta},n_{\xi})=(233,17)$ at $R=76000~a_0$. The percent difference between the two quantities, defined as $200\times(\alpha^{-2}(\xi)-k_{\mathrm{L}}(\xi))/(\alpha^{-2}(\xi)+k_{\mathrm{L}}(\xi))$, is shown as a dotted line. The two vertical lines bound the region where the left hand side of Eq.~\ref{eq:WKBcond} is less than or equal to 0.1.} 
\label{fig:comparekxi}
\end{figure}}

\section{Conclusions}
For low-lying $^2\Sigma$ states of H$_2^+$, we have compared our results with those of other researchers and found that additional refinements near the barrier top are most necessary for the lowest two states. 

For high-lying $^2\Sigma$ states of H$_2^+$ a straightforward WKB approximation yields energies, spectroscopic constants, and wave functions that agree quite well with the quantum results. This level of agreement is not surprising, given that the Langer-corrected WKB energies for non-relativistic hydrogen are exact. Away from the barrier, the quantum mechanical and semiclassical $\nu_2$ values agree to within three or four decimal places. Increased errors due to the barrier arise over a relatively small range of $R$ away from potential wells (if present). The relatively accurate energies away from the barrier top allow one to determine spectroscopic constants that, for the states presented, differ by less than 0.3\% from those obtained by using the quantum mechanical PECs. 

Dipole matrix elements with the ground state also showed good agreement, and were within 30\%, sometimes as close as 2\%. The discrepancies can be attributed to inaccuracies in the wave functions for small $\xi$ or $\eta$, where the WKB assumptions are not valid. The semiclassical phases in these regions are larger than the phases obtained from Milne's phase-amplitude procedure. The states with the biggest discrepancies in phase were those with very few nodes in $\xi$, for which the classical turning points of the potential are close together. These states also show the most differences in their dipole matrix elements. 

It is of course evident that when accurate results are desired, it is preferable to accurately solve the separated angular and radial spheroidal differential equations numerically.  However, this can be time-consuming when many states are needed or a large parameter space needs to be explored.  Then a WKB implementation will often be adequate for many purposes, with a resulting decrease in computational time by about two orders of magnitude. On average, our quantum calculations took 8 minutes on one processor to determine energies of a state within a range of about 2--3\% of $R$, as compared to 2 seconds for the semiclassical calculations.

\begin{acknowledgement}

The authors thank Jacob Covey and Manuel Endres for stimulating discussions that led us to the current exploration.  This work was supported in part by the U.S. Department of Energy, Office of Science, under Award No. DE-SC0010545.

\end{acknowledgement}

\clearpage
\newpage

\bibliography{h2plusbib}

\end{document}